\documentclass[final,5p,times,twocolumn,authoryear]{elsarticle}
\usepackage{amssymb}
\usepackage{graphicx}
\usepackage{array}
\usepackage{hyperref}

\usepackage{xcolor}
\usepackage{ulem}

\journal{Journal of High Energy Astrophysics}
\begin{document}
\begin{frontmatter}

\title{Ultraviolet variability in Radio-Loud Active Galactic Nuclei observed by UVIT onboard \textit{AstroSat}}

\author[first]{M. Reshma}
\author[second]{C. S. Stalin}
\author[third]{Amit Kumar Mandal}
\author[first]{S. B. Gudennavar \thanks{E-mail: shivappa.b.gudennavar@christuniversity.in}}
\author[fourth]{Senorita Benedict}
\author[second]{Prajwel Joseph}

\affiliation[first]{organization={Department of Physics and Electronics, CHRIST University}, 
            city={Bengaluru 560029}, state={Karnataka}, country={India}}

\affiliation[second]{organization={Indian Institute of Astrophysics}, 
            addressline={Block II, Koramangala}, city={Bengaluru 560034}, state={Karnataka}, country={India}}

\affiliation[third]{organization={Center for Theoretical Physics of the Polish Academy of Sciences}, 
            addressline={Al. Lotniko´w 32/46}, city={02–668 Warsaw}, country={Poland}}
            
\affiliation[fourth]{organization={St. Joseph's University}, 
            addressline={36, Lal Bagh Main Road, Langford Gardens}, city={Bengaluru 560 027}, country={India}}
            
\begin{abstract}
Radio-loud active galactic nuclei (AGN) are among the most luminous objects in the Universe, emitting radiation from low-energy radio waves to high energy $\gamma$-rays. They are well known to exhibit flux variations at nearly all  accessible wavelengths. However, their variability properties in the ultraviolet (UV) band remain relatively less explored compared to other wavebands. Here, we present  the results of a systematic  investigation of the  UV flux and spectral variability characteristics of 24 radio-loud AGN spanning the redshift range 0.018 $\le$ $z$ $\le$ 2.218. The sample comprises 17 BL Lac objects, 6 flat spectrum radio quasars (FSRQs) and one radio-loud narrow line Seyfert 1 galaxy. We used observations obtained with the Ultra-Violet Imaging Telescope (UVIT) onboard {\it AstroSat} during its first ten years of operation, covering both the far-UV (FUV; 1300$-$1800 \AA) and near-UV (NUV; 2000$-$3000 \AA) bands. Of the 24 sources analysed, 18 showed significant UV variability  on  hour timescales. We found a bluer when brighter (BWB) spectral trend in two sources: the FSRQ CTA 102  and the BL Lac PKS 0447$-$439. The observed UV variability in our sample of radio-loud AGN, together with the BWB trend detected in these two sources, supports a scenario in which the hour timescale UV variations are driven by intrinsic processes within their relativistic jets. 
\end{abstract}

\end{frontmatter}

\section{Introduction}
\label{introduction}

Active galactic nuclei (AGN) are among the most luminous sources in the Universe, with luminosities in the range 10$^{11}$ $-$ 10$^{14}$ L$_{\odot}$. Such extreme emission is powered by the accretion of matter onto supermassive black holes (SMBHs) located at the centres of galaxies, with masses spanning $10^6$ to $10^{10}$ M$_\odot$  \citep{lynden1969galactic,rees1984black}. The accretion process that powers AGN produces radiation across the entire electromagnetic spectrum from low-energy radio waves to high-energy $\gamma$-rays. AGN  are broadly classified into two main categories based on their radio properties: radio-loud AGN, which emit predominantly  non-thermal radiation through powerful relativistic jets, and radio-quiet AGN, which lack prominent radio jets and whose emission is dominated by thermal radiation from their accretion disks. The fundamental distinction between these two classes lies in the presence or absence of strong relativistic jets. Approximately 15\% of AGN are radio-loud \citep{kellermann1989vla}. This population includes radio galaxies, blazars, and radio-loud narrow-line Seyfert 1 galaxies. However, it has been suggested that AGN need to be classified based on a physical difference into non-jetted (absence of relativistic jets; eg. Seyferts) and jetted (presence of relativistic jets; eg. blazars) AGN \citep{padovani2017active}. 
\\[6pt]
Blazars constitute a prominent subclass of radio-loud AGN, and are  characterized by relativistic jets oriented closely to the observer’s line of sight, typically within a viewing angle $\le 10^\circ$ \citep{urry1995unified}. These sources exhibit rapid and large amplitude flux variability on timescales ranging from minutes to hours \citep{wagner1995intraday,ulrich1997variability,raiteri2025variability}, spanning the entire electromagnetic spectrum from radio to $\gamma$-rays. In the high energy $\gamma$-ray band, flux variations on minute timescales have been detected in a handful of blazars \citep{aharonian2007exceptional,albert2007variable,arlen2013rapid,ackermann2016minute,shukla2018short,shukla2020gamma,pandey2022detection}. Blazars are further classified into two subclasses: flat spectrum radio quasars (FSRQs), which show broad emission lines with equivalent widths $>$ 5 \AA, and BL Lacertae objects (BL Lacs), which show absent or weak  emission lines with equivalent widths $<$ 5 \AA\ \citep{urry1995unified}. However, according to \cite{ghisellini2011transition}, the fundamental physical distinction between FSRQs and BL Lacs is based on the broad-line region (BLR) luminosity relative to the Eddington luminosity, where FSRQs are characterized by a luminous BLR satisfying $\rm L_{BLR}/L_{Edd}$ $>$ 5 $\times$ $10^{-4}$. Their broadband spectral energy distribution (SED) typically displays a characteristic double-peaked structure. The low-energy peak, extending from the infrared to the ultraviolet (UV)/soft X-ray band, is attributed to synchrotron emission from relativistic electrons in the jet. The high-energy peak, occurring at the hard X-ray/$\gamma$-ray region, is commonly interpreted as arising from inverse Compton scattering of synchrotron photons (synchrotron self-Compton) or external photons (external Compton). Based on the frequency of the synchrotron peak, blazars are further categorized as low-synchrotron-peaked (LSP; $\nu_{peak}$ $<$ $10^{14}$ Hz), intermediate-synchrotron-peaked (ISP; $10^{14}$ $<$ $\nu_{peak}$ $<$ $10^{15}$ Hz) and high-synchrotron-peaked (HSP; $\nu_{peak}$ $>$ $10^{15}$ Hz) sources \citep{abdo2010spectral}.
\\[6pt]
Another important class of  AGN is the  narrow-line Seyfert 1  (NLSy1) galaxies, first identified by \cite{osterbrock1985spectra}. These sources are characterised by narrow H$\beta$ emission lines with full width at half maximum $<$ 2000 km s$^{-1}$, [OIII]/H$\beta$ $<$ 3, prominent FeII emission, steep soft X-ray spectra, and rapid X-ray variability \citep{boller1996soft,leighly1999comprehensive,1999ApJS..125..297L}. NLSy1 galaxies are thought to host relatively  low-mass black holes (10$^6$ $-$ 10$^8$ M$_\odot$; \citealt{peterson2000x,grupe2004mbh}). Approximately 5$-$7\% of NLSy1 galaxies are radio-loud \citep{komossa2006radio,yuan2008population,rakshit2017catalog}. Unlike blazars, which are typically hosted by elliptical galaxies, radio-loud NLSy1 galaxies are believed to reside in spiral galaxies \citep{belloni2010jet}, with some exhibiting large scale relativistic jets \citep{rakshit2018rare,vietri2022spectacular,umayal2025identification}. A small fraction of radio-loud NLSy1 galaxies are detected as high energy $\gamma$-ray emitters by the Fermi-Large Area Telescope \citep{abdo2009radio,paliya2018gamma} pointing to the presence of relativistic jets in them \citep{d2019relativistic,foschini2020jetted}. They exhibit intra-night optical variability comparable to that observed in blazars \citep{paliya2013intranight}. Rapid variability in X-rays, UV and optical bands are also known in these objects \citep{d2020gamma,d2020short}. These findings suggest that the jet contributes to the observed optical and UV emission, while in X-rays, the emission likely arises from both the jet and corona. There is also a report of an NLSy1 galaxy exhibiting rapid X-ray variability, with a halving timescale of less than 30 seconds \citep{feigelson1986h}.
\\[6pt]
Variability is a defining observational property of AGN \citep{wagner1995intraday,ulrich1997variability,raiteri2025variability} and serves as a powerful diagnostic for probing the physical processes governing accretion and jet emission. The dominant emission mechanism in the UV band depends on both the source type and its activity state. In radio-quiet (non-jetted) AGN, the UV emission primarily arises from thermal radiation emitted by the accretion disk, typically manifested as the “big blue bump” in their broadband SED. In contrast, in radio-loud (jetted) AGN, the UV emission is largely dominated by synchrotron radiation originating from their relativistic jets. In the broadband SEDs of blazars, the UV emission is predominantly produced  by synchrotron radiation from relativistic electrons in the jet \citep{abdo2011fermi,paliya2015multi}. However, in FSRQs, during low-flux states, signatures of accretion disk emission are seen in their broadband SEDs \citep{paliya2016broadband,paliya2017general}. Despite its importance, systematic studies of AGN  variability in the UV regime remain relatively scarce \citep{paltani1994systematic,welsh2011galex, sukanya2018long, reshma2024ultraviolet}. 
Using observations from the International Ultraviolet Explorer (IUE), \cite{paltani1994systematic}  reported increasing variability amplitude  toward shorter wavelengths for a large sample of AGN,  while \cite{edelson1991rapid} detected rapid UV variability on timescales shorter than a day in the blazar PKS 2155$-$304.  UV variability nature of AGN is therefore less understood.
\\[6pt]
In this work, we investigated the UV flux and spectral variability on hour timescales for a sample of 24 AGN, comprising one radio-loud NLSy1 galaxy, 17 BL Lacs, and 6 FSRQs using observations obtained with the Ultra-Violet Imaging Telescope (UVIT; \citealt{tandon2020additional}) onboard \textit{AstroSat}. Our sample selection was guided by two criteria: (a) the source must be radio-loud, and (b) the observational data must be publicly available beyond the proprietary period. Applying these criteria yielded a sample of 24 AGN. All are blazars, with the exception of 1H 0323+342, which is a radio-loud NLSy1 galaxy. However, 1H 0323+342 shares several properties with flat-spectrum quasars: it has been detected in the $\gamma$-ray band \citep{abdo2009radio}, and its broadband SED closely resembles that of FSRQs. For these reasons, we have included it in our sample. The structure of the paper is as follows: Section \ref{sec:Observations and data reduction} describes the data collection and data reduction procedures. Section \ref{sec:Analysis} outlines the methods  employed for flux and spectral variability analysis. Notes on individual sources are given in Section \ref{sec:Individual source description}. Results and discussion are given in Section \ref{sec:discussion}, followed by the summary in the final Section \ref{sec:summary}.

\section{Observations and data reduction}
\label{sec:Observations and data reduction}

\begin{figure*}
\hbox{
\hspace{0.5cm}
\includegraphics[scale=0.28]{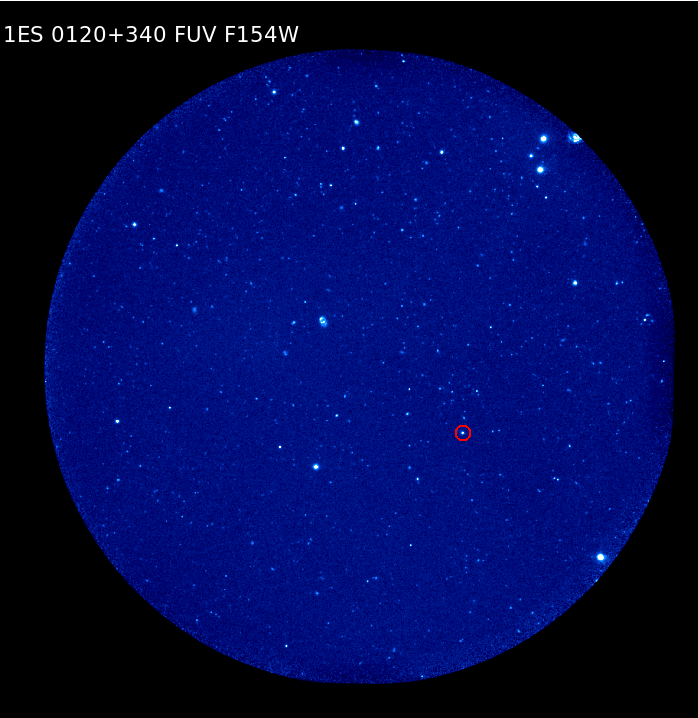}
\hspace{0.1cm}
\includegraphics[scale=0.28]{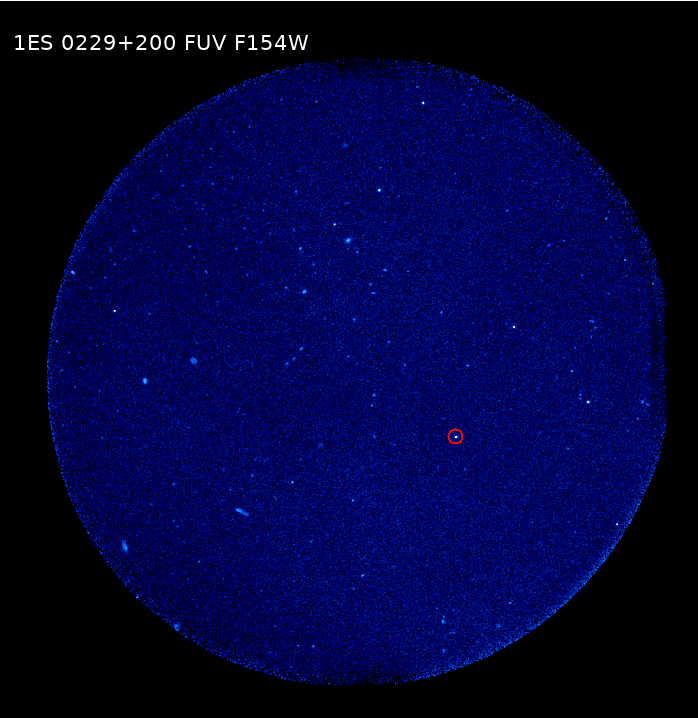}
\hspace{0.1cm}
\includegraphics[scale=0.28]{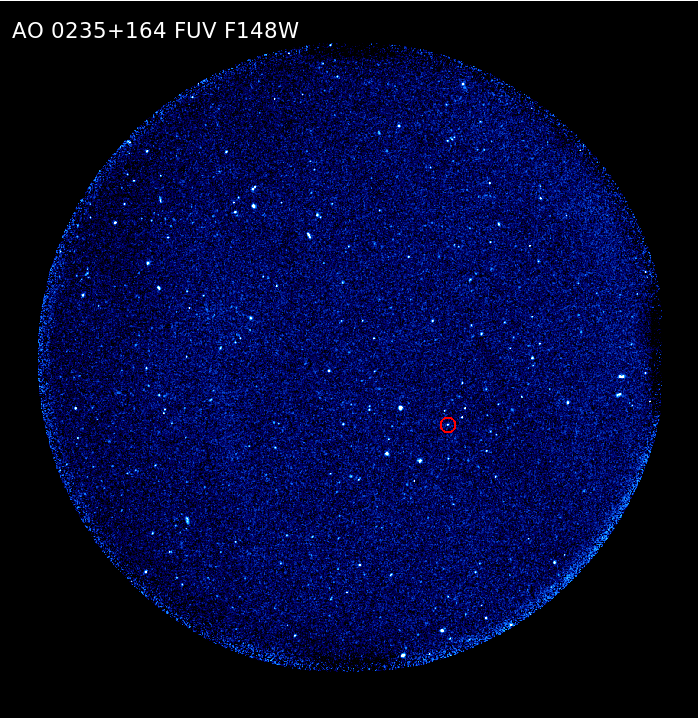}
}
\vspace{0.15cm}
\hbox{
\hspace{0.5cm}
\includegraphics[scale=0.28]{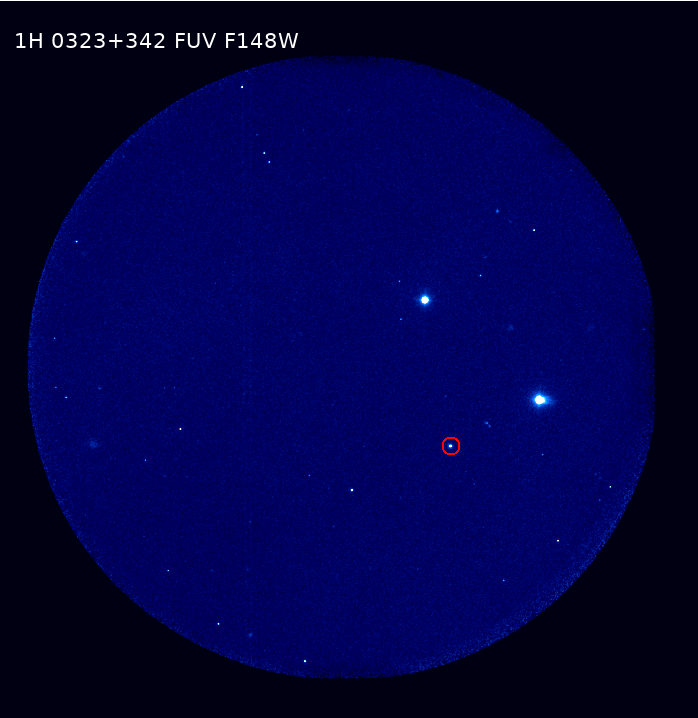}
\hspace{0.1cm}
\includegraphics[scale=0.28]{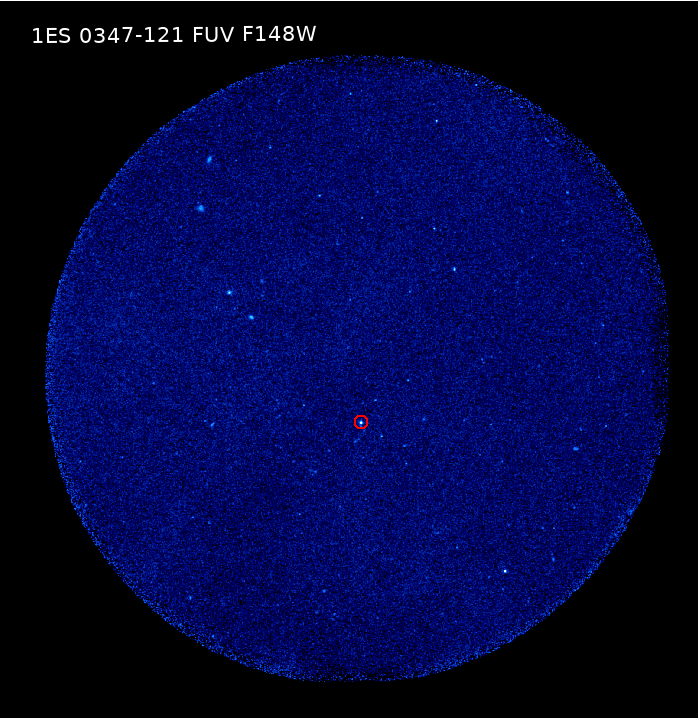}
\hspace{0.1cm}
\includegraphics[scale=0.28]{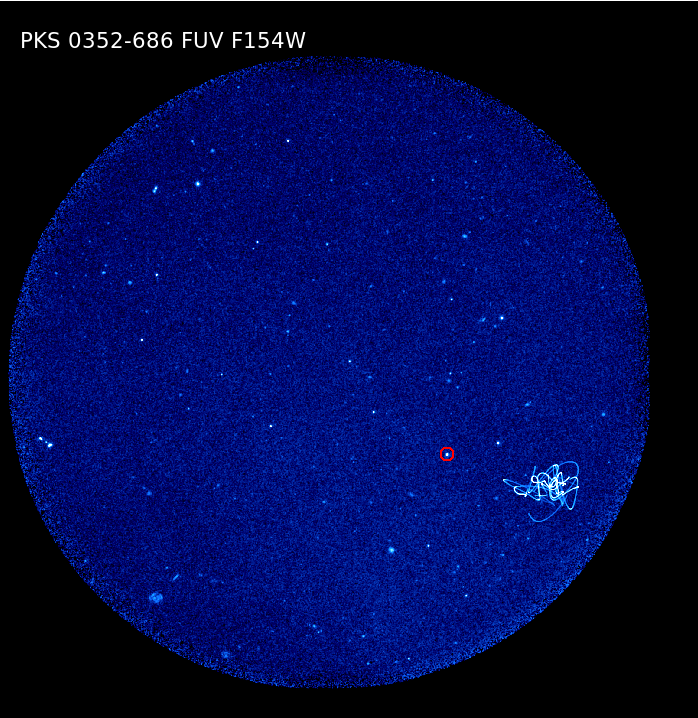}
}
\vspace{0.15cm}
\hbox{
\hspace{0.5cm}
\includegraphics[scale=0.28]{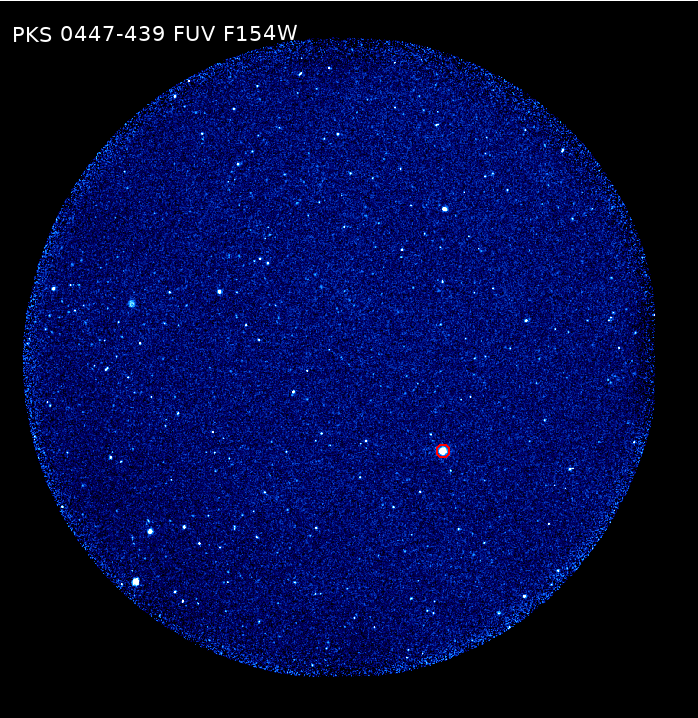}
\hspace{0.1cm}
\includegraphics[scale=0.28]{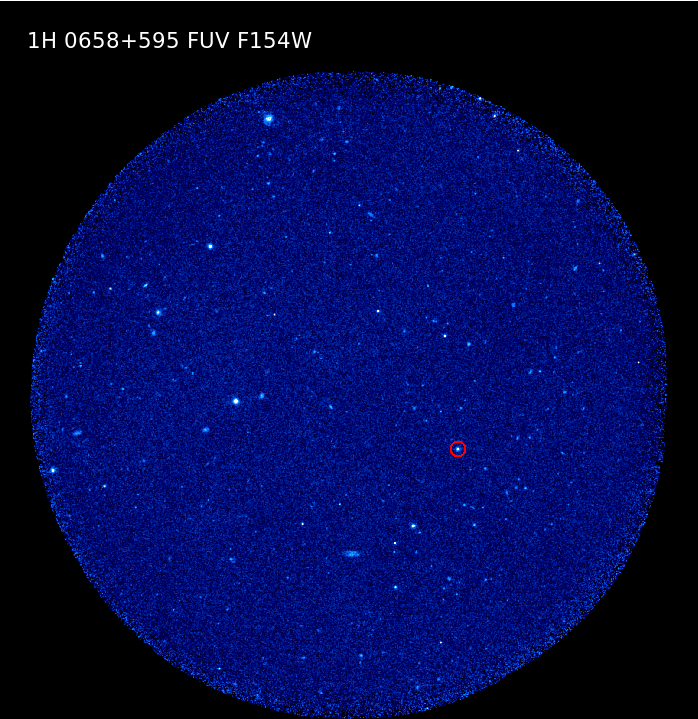}
\hspace{0.1cm}
\includegraphics[scale=0.28]{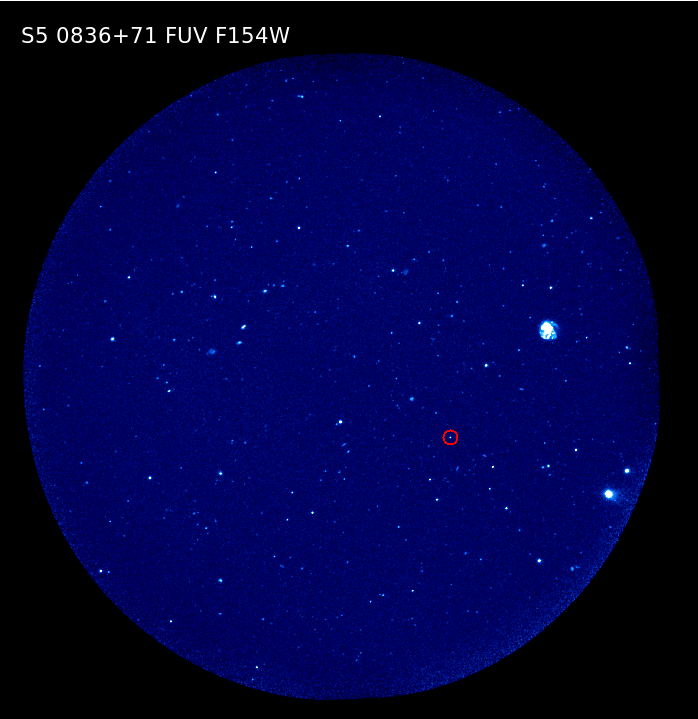}
}
\vspace{0.15cm}
\hbox{
\hspace{0.5cm}
\includegraphics[scale=0.28]{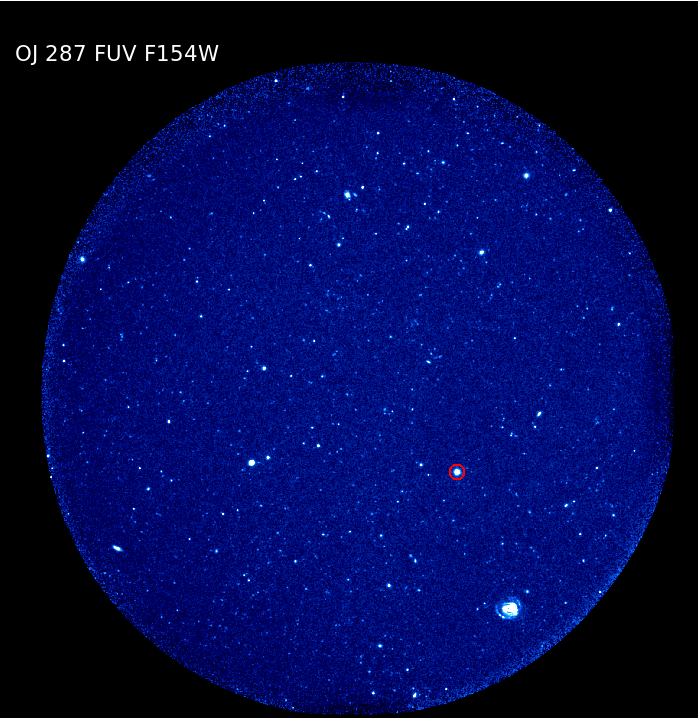}
\hspace{0.1cm}
\includegraphics[scale=0.28]{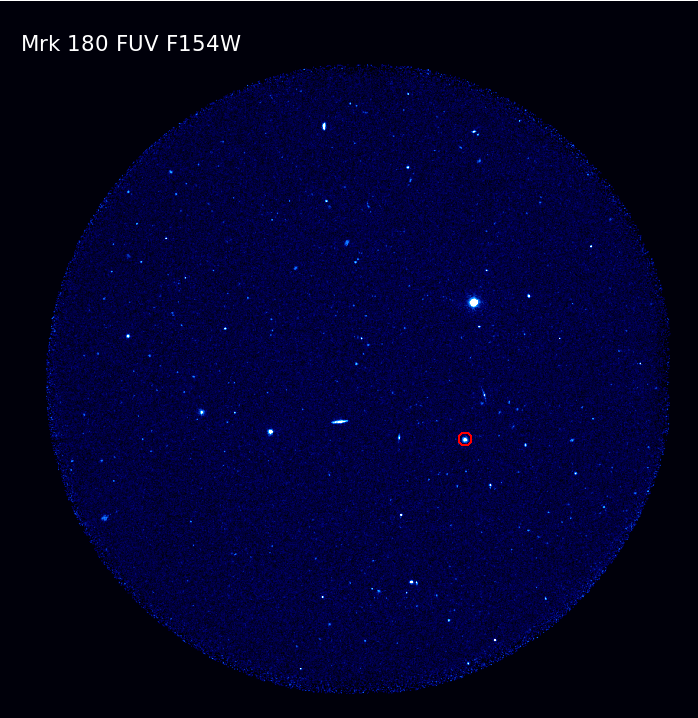}
\hspace{0.1cm}
\includegraphics[scale=0.28]{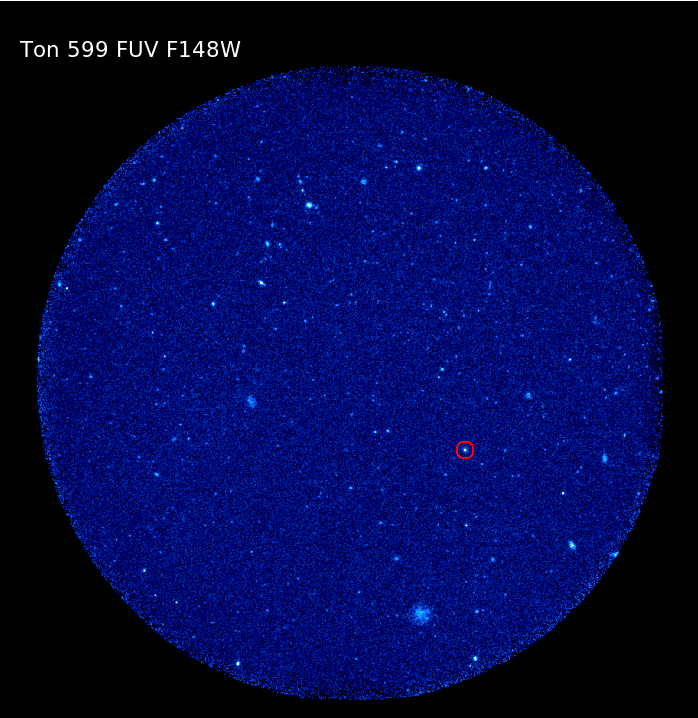}
}
\caption{FUV field images of the blazar sources analyzed in this work. Red circles in each field image shows the target sources. The source name and the respective filter name are shown in all the field images. The field of view of each image is $\sim$ $28^\prime$ diameter. In the field image of PKS 0352$-$686, the pattern observed near the target source is due to an unmasked hot pixel showing the drift correction patterns.}
\label{figure-1}
\end{figure*}

\begin{figure*}
\vspace{0.15cm}
\hbox{
\hspace{0.5cm}
\includegraphics[scale=0.28]{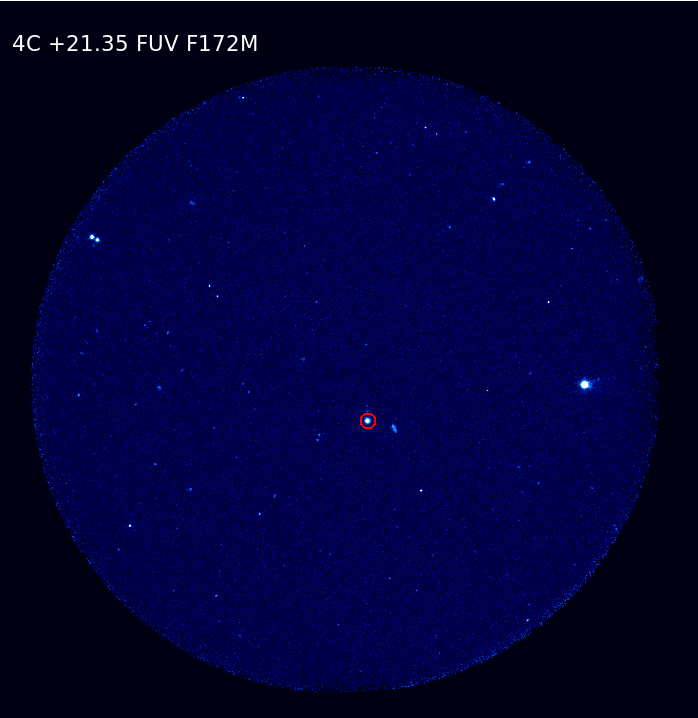}
\hspace{0.1cm}
\includegraphics[scale=0.28]{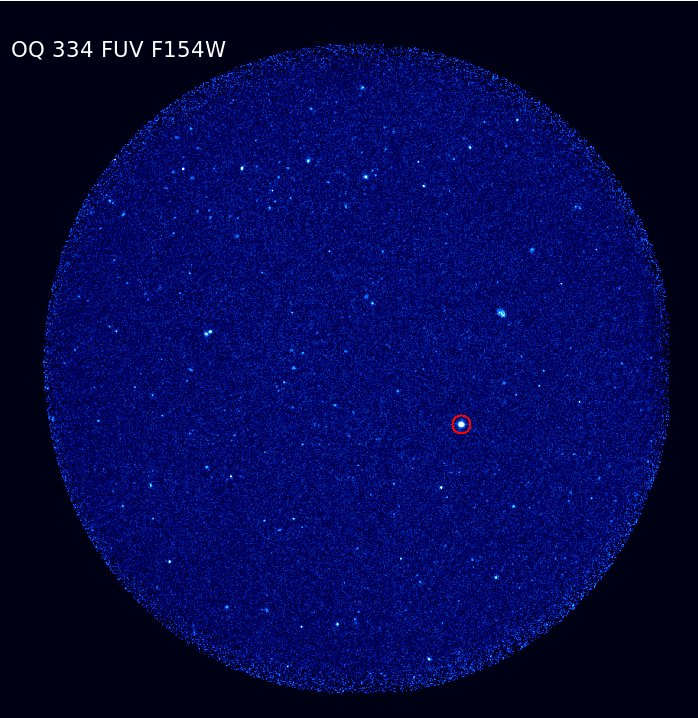}
\hspace{0.1cm}
\includegraphics[scale=0.28]{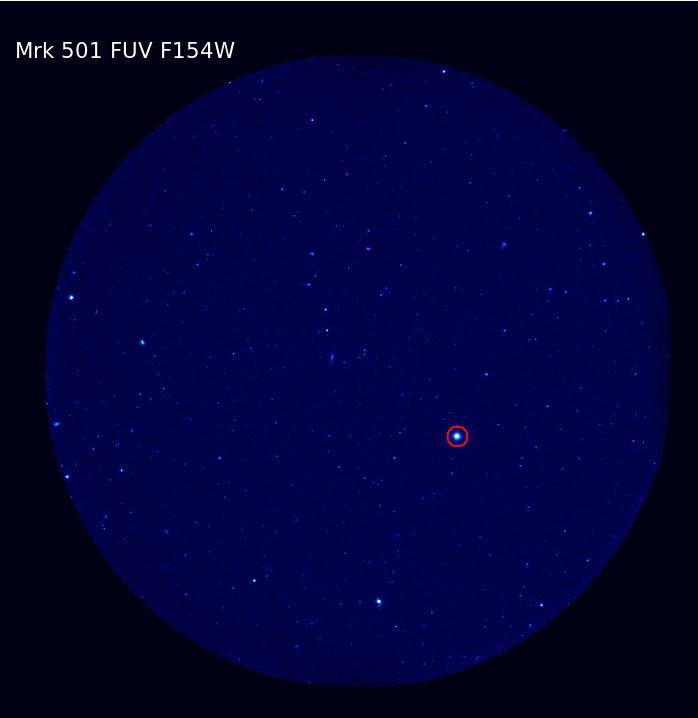}
}
\vspace{0.15cm}
\hbox{
\hspace{0.5cm}
\includegraphics[scale=0.28]{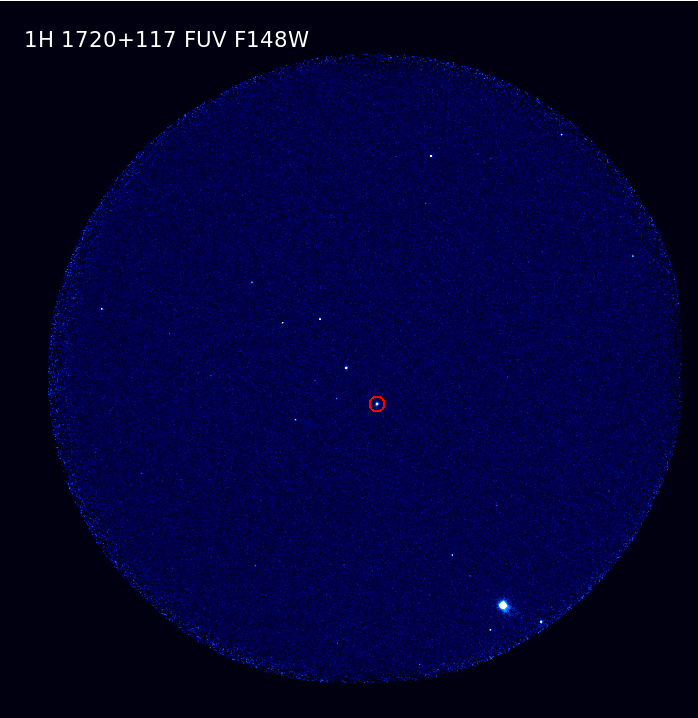}
\hspace{0.1cm}
\includegraphics[scale=0.28]{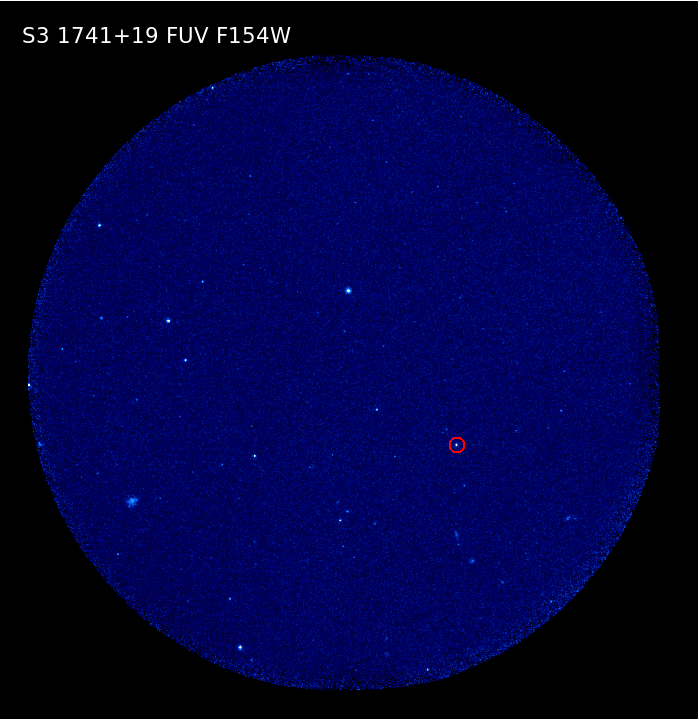}
\hspace{0.1cm}
\includegraphics[scale=0.28]{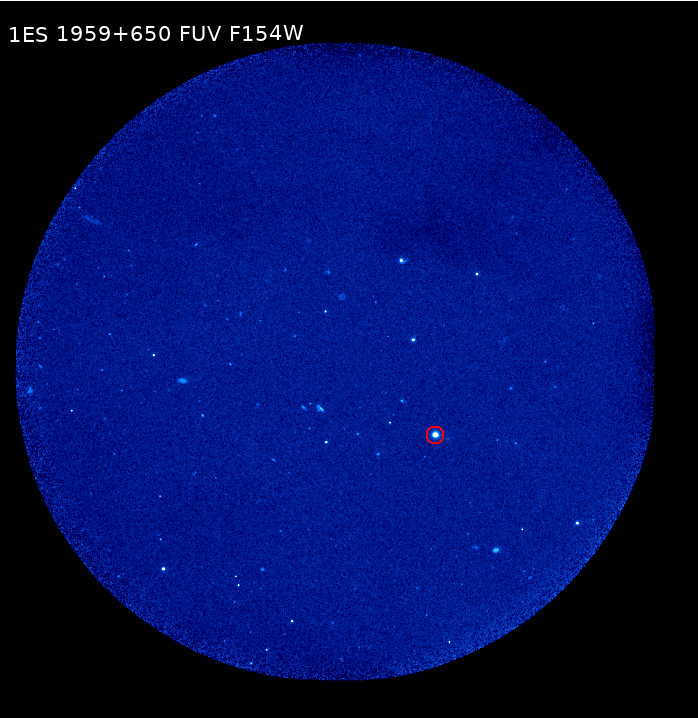}
}
\vspace{0.15cm}
\hbox{
\hspace{0.5cm}
\includegraphics[scale=0.28]{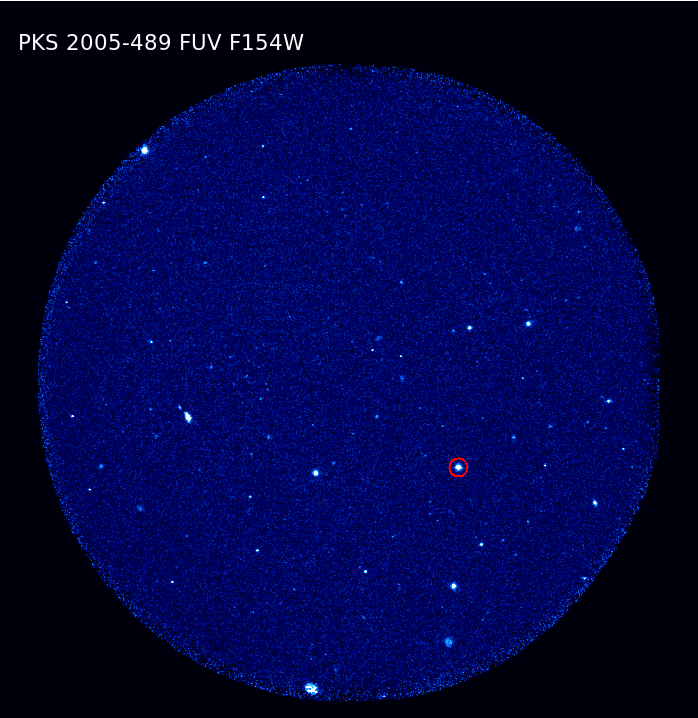}
\hspace{0.1cm}
\includegraphics[scale=0.28]{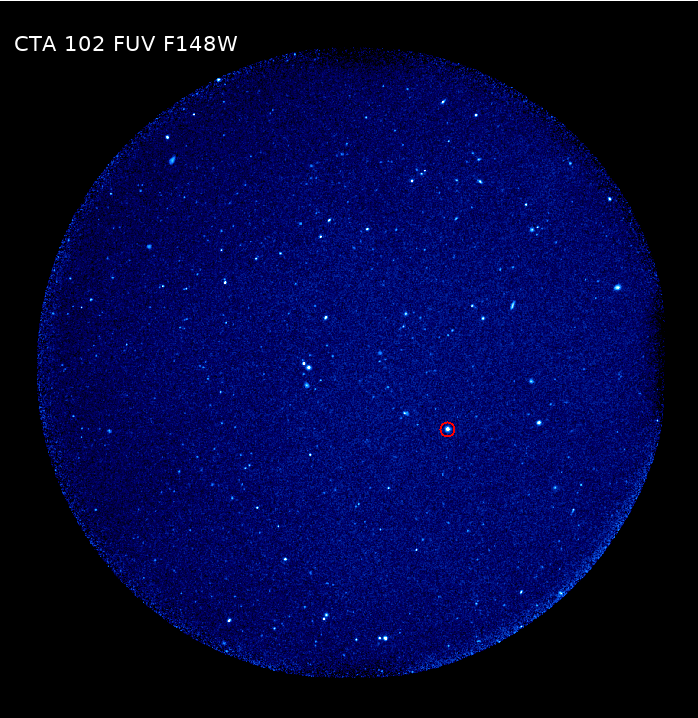}
\hspace{0.1cm}
\includegraphics[scale=0.28]{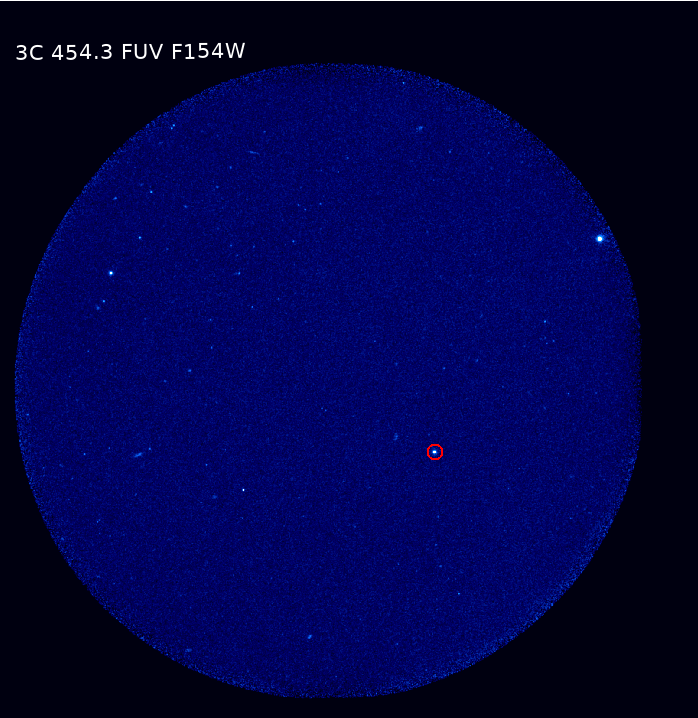}
}
\vspace{0.15cm}
\hbox{
\hspace{0.5cm}
\includegraphics[scale=0.28]{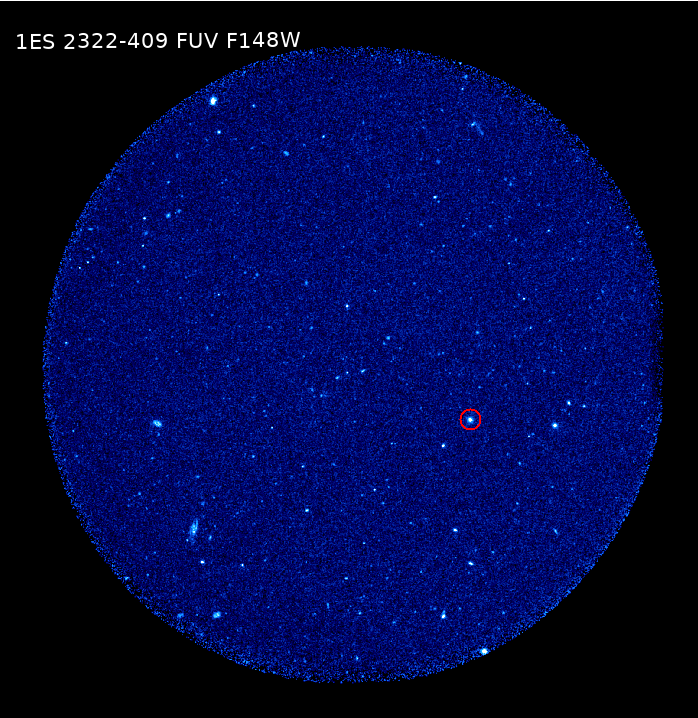}
\hspace{0.1cm}
\includegraphics[scale=0.28]{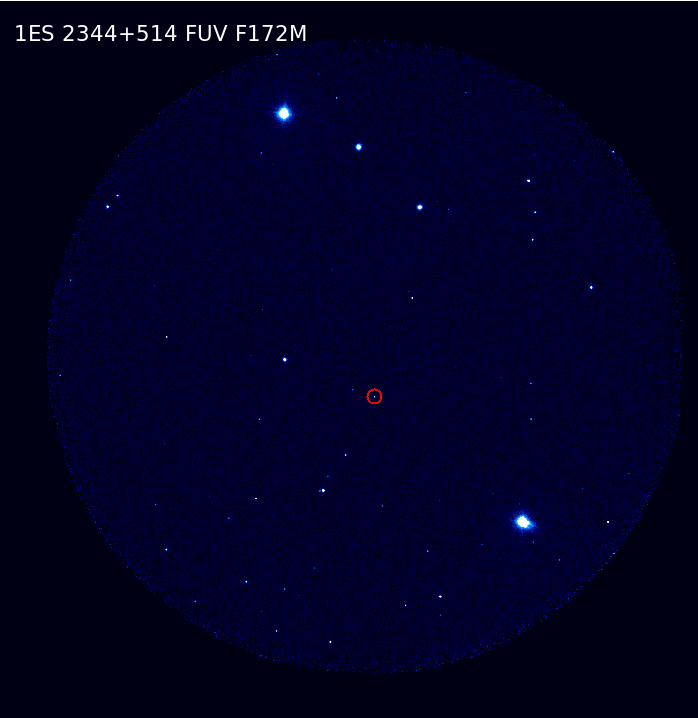}
\hspace{0.1cm}
\includegraphics[scale=0.28]{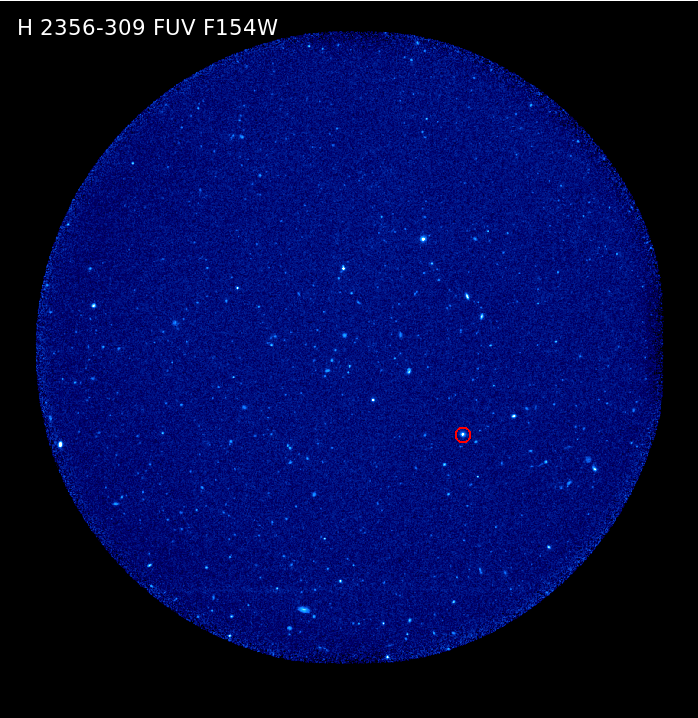}
}
\caption{FUV field images of the blazar sources analyzed in this work. Red circles in each field image shows the target sources. The source name and the respective filter name are shown in all the field images. The field of view of each image is $\sim$ $28^\prime$ diameter.}
\label{figure-2}
\end{figure*}

We utilized publicly available archival observations of radio-loud (jetted) AGN obtained with UVIT onboard \textit{AstroSat} \citep{agrawal2017astrosat}, during the first ten years of its operation following its launch by the Indian Space Research Organization, Bengaluru on 28 September 2015. UVIT has  a circular field of view of approximately 28$'$ in diameter and provides simultaneous imaging in the far-UV (FUV; 1300$-$1800 \AA) and near-UV (NUV; 2000$-$3000 \AA) bands. It also includes a visible channel (VIS; 3200$-$5500 \AA),  which is primarily used for tracking the telescope aspect.  Table \ref{Table-1} lists the sources analysed in this work, while Table \ref{Table-2} provides the corresponding observation log. 
\\[6pt]
We used the Level 1 data obtained from the Indian Space Science Data Center (ISSDC) to generate the science ready images. These images were generated by the UVIT Payload Operation Center (POC) at the Indian Institute of Astrophysics, Bengaluru, using version 7.0 of the UVIT Level 2 data processing pipeline \citep{joseph2025uvit}. The science ready images made available to ISSDC for archival and dissemination by the POC were created from the raw Level 1 data after applying corrections for spacecraft drift, flat field effects and geometric distortions. The UVIT observed fields of the sources are shown in Figs. \ref{figure-1} and \ref{figure-2}. To investigate UV flux variability, we used the Curvit Python package \citep{joseph2021curvit}. We generated orbit-wise lightcurves from the combined Level 2 events list using the \textit{curvit} function \textit{curve$\_$orbitwise()}, by extracting counts from a circular source region of radius 12 sub-pixels ($\sim$ $5^{\prime\prime}$). The extracted lightcurves were corrected for background, aperture effects, and saturation. The resulting fluxes were converted to the total brightness values using  the conversion factors provided in Table 11 of \cite{tandon2020additional}. In each light curve, we excluded photometric data points with uncertainties greater than twice the median error of all points in that light curve. The lightcurves are presented in Figs. \ref{figure-3}, \ref{figure-4} and \ref{figure-5}. The observed UV brightness values (not corrected for galactic extinction) of all the  sources are given in Appendix. 

\begin{table*}
\centering
\caption{Information of the sources studied in this work. The details of the sources are from \cite{ackermann2015third} except for the redshift of 1H 1720+117 which is from  \href{url}{https://simbad.u-strasbg.fr/simbad/sim-fbasic}. All the sources are detected in $\gamma$-rays by \textit{Fermi}-LAT.}
\begin{tabular}{lccccccl}
&&&&&\\
\hline
Name & RA (2000)        & Dec (2000)      & Optical & SED  & $z$  & $M_{BH}$  & References  \\
     & (hh:mm:ss)       & (dd:mm:ss)      & type    & type &  & ($M_\odot$)  & for $M_{BH}$ \\ 
\hline
1ES 0120+340    & 01:23:08.55 & +34:20:47.40   & BL Lac & HSP & 0.272 & 1.0 $\times 10^{9}$     &  \cite{goswami2024variety}   \\
1ES 0229+200    & 02:32:48.61 & +20:17:17.47   & BL Lac & HSP & 0.139 & 4.8 $\times 10^{8}$     &  \cite{woo2005black}       \\
AO 0235+164     & 02:38:38.93 & +16:36:59.27   & BL Lac     & LSP & 0.940 & (4.7 $\pm$ 2.0) $\times 10^{8}$     &  \cite{liu2006harmonic}     \\
1H 0323+342     & 03:24:41.16 & +34:10:45.85   & NLSy1  & HSP & 0.063 & (2.8 -7.9) $\times 10^{6}$     &  \cite{pan2018independent}     \\
1ES 0347$-$121  & 03:49:23.17 & $-$11:59:27.20 & BL Lac     & HSP & 0.188 & 1.0 $\times 10^{8}$     &   \cite{woo2005black}    \\
PKS 0352$-$686  & 03:52:57.01 & $-$68:31:18.59 & BL Lac     & HSP & 0.087 &  -    &    -   \\
PKS 0447$-$439  & 04:49:24.69 & $-$43:50:08.96 & BL Lac     & HSP & 0.205 & 6.0 $\times 10^{8}$  & \cite{ghisellini2010general}   \\ 
1H 0658+595     & 07:10:30.06 & +59:08:20.36   & BL Lac     & HSP & 0.125 & 1.8 $\times 10^{8}$     & \cite{woo2005black}     \\
S5 0836+71      & 08:41:24.36 & +70:53:42.17   & FSRQ   & LSP & 2.218 & 5.0 $\times 10^{9}$ &  \cite{tagliaferri2015nustar}    \\
OJ 287          & 08:54:48.87 & +20:06:30.64   & BL Lac     & LSP & 0.306 & 1.8 $\times 10^{10}$ & \cite{kuznetsov2024black}      \\ 
Mrk 180         & 11:36:26.40 & +70:09:27.30   & BL Lac     & HSP & 0.045 & 1.7 $\times 10^{8}$ & \cite{woo2005black} \\
Ton 599         & 11:59:31.83 & +29:14:43.82   & FSRQ   & LSP & 0.725 & 1.0 $\times 10^{8}$  &  \cite{maurya2025origin}    \\ 
4C +21.35       & 12:24:54.45 & +21:22:46.38   & FSRQ   & LSP & 0.435 & 6.0 $\times 10^{8}$  &  \cite{farina2012optical}    \\
OQ 334         & 14:22:30.37 & +32:23:10.44   & FSRQ   & LSP & 0.682 & 3.9 $\times 10^{8}$ &   \cite{prince2021broad}     \\
Mrk 501        & 16:53:52.21 & +39:45:36.60   & BL Lac     & HSP & 0.034 & 4.2 $\times 10^{8}$ &  \cite{woo2005black}     \\
1H 1720+117    & 17:25:04.34 & +11:52:15.47   & BL Lac     & HSP & 0.018 &  -    &    -   \\ 
S3 1741+19     & 17:43:57.83 & +19:35:09.01   & BL Lac     & HSP & 0.084 & 1.0 $\times 10^{9}$  &  \cite{goswami2024variety}    \\
1ES 1959+650   & 19:59:59.85 & +65:08:54.65   & BL Lac     & HSP & 0.047 & 2.0 $\times 10^{8}$ &  \cite{ghisellini2010general}    \\
PKS 2005$-$489 & 20:09:25.39 & $-$48:49:53.72 & BL Lac     & HSP & 0.071 & 5.0 $\times 10^{8}$ &   \cite{ghisellini2010general}   \\
CTA 102        & 22:32:36.40 & +11:43:50.90   & FSRQ   & LSP & 1.037 & 8.5 $\times 10^{8}$ & \cite{sahakyan2020investigation}      \\
3C 454.3       & 22:53:57.74 & +16:08:53.56   & FSRQ   & LSP & 0.859  & 1.5 $\times 10^{9}$ & \cite{sahakyan2021modelling} \\
1ES 2322$-$409 & 23:24:44.89 & $-$40:40:43.89 & BL Lac     & HSP & 0.174 &  1.0 $\times 10^{9}$ &  \cite{goswami2024variety}     \\
1ES 2344+514   & 23:47:04.83 & +51:42:17.88   & BL Lac     & HSP & 0.044 & 5.5 $\times 10^{8}$ & \cite{woo2005black}\\
H 2356$-$309   & 23:59:07.90 & $-$30:37:40.67   & BL Lac    & HSP & 0.165 & 1.5 $\times 10^{8}$ &  \cite{woo2005black}    \\
\hline
\end{tabular}
\label{Table-1}
\end{table*}

\begin{table*}
\centering
\caption{Log of observations. Here OBSID: Observation ID and MJD: Modified Julian Date}
\begin{tabular}{lcccccc}
 & &&&&& \\
\hline 
Name & Date of observation  & OBSID & Filter & MJD  & MJD  & Net exposure  \\
     & (dd-mm-yyyy) &     &       & Start & End  &  time (s)\\
\hline
1ES 0120+340   & 04-12-2018 & A05\_185T04\_9000002548 & F154W & 58453.058824  & 58456.581583  & 64723  \\
1ES 0229+200   & 22-09-2017 & A03\_078T01\_9000001546 & F154W & 58017.831844  & 58018.174450  & 4889   \\
               &            &                       & F172M & 58018.176239  & 58018.674273  & 8724   \\
               &            &                       & N219M & 58017.831844  & 58018.246042  & 5787   \\
               &            &                       & N245M & 58018.310158  & 58018.467429  & 3703   \\
               &            &                       & N263M & 58018.469098  & 58018.674346  & 3453   \\
               & 09-08-2021 & T04\_034T01\_9000004632 & F154W & 59434.458958  & 59435.481634  & 14007  \\ 
               &            &                       & F169M & 59435.483371  & 59436.773205  & 15032  \\
AO 0235+164     & 04-09-2016 & G05\_218T04\_9000000640 & F148W & 57634.275091  & 57635.847420  & 22651  \\ 
               &            &                       & N219M & 57634.275042  & 57635.847469  & 22637  \\
1H 0323+342    & 06-09-2020 & T03\_223T01\_9000003850 & F148W & 59098.402173  & 59098.757547  & 8089   \\  
               & 30-09-2020 & T03\_236T01\_9000003908 & F148W & 59122.499163  & 59122.765409  & 6579   \\
               & 04-02-2022 & A05\_019T01\_9000004906 & F148W & 59613.668914  & 59614.422499  & 9406   \\
1ES 0347$-$121 & 21-08-2020 & A07\_146T01\_9000003828 & F148W & 59082.304138  & 59082.519524  & 2942   \\
PKS 0352$-$686 & 02-11-2018 & A05\_202T01\_9000002484 & F148W & 58423.476028  & 58423.896043  & 11368  \\
               &            &                       & F154W & 58423.897781  & 58424.569960  & 11415  \\
PKS 0447$-$439 & 15-12-2017 & A04\_082T01\_9000001772 & F154W & 58102.012648  & 58102.544877  & 9956   \\
               &            &                       & F169M & 58102.546607  & 58102.562569  & 1360   \\
               &            &                       & N245M & 58102.012599  & 58102.544927  & 10031  \\
               &            &                       & N263M & 58102.546558  & 58102.562619  & 1367   \\
1H 0658+595  & 19-11-2016 & A02\_085T02\_9000000808 & F154W & 57711.076889  & 57711.427463  & 5556   \\
               &            &                       & F172M & 57711.440560  & 57711.706765  & 5995   \\
               &            &                       & N245M & 57711.437081  & 57711.571027  & 2874   \\
               &            &                       & N263M & 57711.577500  & 57711.588043  & 870    \\
               &            &                       & N279N & 57711.076839  & 57711.432454  & 4756   \\
S5 0836+71     & 04-05-2017 & G07\_024T01\_9000001206 & F154W & 57875.583540  & 57878.496745  & 58336  \\
OJ 287         & 25-10-2020 & T03\_249T01\_9000003934 & F154W & 59142.523770  & 59151.388587  & 52665  \\  
               & 18-11-2021 & A10\_067T01\_9000004770 & F154W & 59535.873750  & 59536.902358  & 12262  \\
               & 20-11-2021 & A10\_067T01\_9000004774 & F154W & 59537.712823  & 59538.863612  & 11921  \\
               & 18-04-2023 & A12\_104T02\_9000005572 & F154W & 60051.025594  & 60052.712404  & 21125  \\ 
Mrk 180        & 25-10-2018 & A05\_064T01\_9000002450 & F148W & 58415.633992  & 58416.584151  & 10233  \\
               &            &                       & F154W & 58416.585873  & 58416.863542  & 4500   \\
Ton 599        & 24-04-2020 & T03\_194T01\_9000003620 & F148W & 58962.537295  & 58962.886370  & 7582   \\
               &            &                       & F169M & 58962.888109  & 58963.557386  & 9358   \\ 
               &            &                       & F172M & 58963.559120  & 58963.905821  & 9197   \\   
4C +21.35       & 25-04-2018 & A04\_219T01\_9000002054 & F148W & 58233.540491  & 58233.609904  & 1902   \\
               &            &                       & F172M & 58233.611642  & 58233.891753  & 7434     \\  
OQ 334         & 12-01-2020 & T03\_175T01\_9000003428 & F148W & 58859.704781  & 58859.979071  & 3823   \\
               &            &                       & F154W & 58859.980809  & 58860.518063  & 3788   \\
               &            &                       & F169M & 58860.519803  & 58860.793514  & 3818   \\
               &            &                       & F172M & 58860.795252  & 58860.798112  & 233    \\ 
Mrk 501        & 25-03-2022 & T05\_015T01\_9000005026 & F154W & 59662.314522  & 59664.496390  & 35246  \\ 
               & 23-10-2022 & A07\_145T01\_9000005370 & F148W & 59870.516163  & 59876.809498  & 19270    \\
1H 1720+117    & 10-05-2018 & A04\_224T05\_9000002088 & F148W & 58248.765601  & 58248.845706  & 2722   \\
S3 1741+19     & 23-08-2019 & A05\_163T01\_9000003118 & F154W & 58717.672011  & 58718.287319  & 7615   \\
               &            &                       & F172M & 58718.289059  & 58718.830209  & 9845   \\ 
\hline
\end{tabular}
\end{table*}

\addtocounter{table}{-1}

\begin{table*}
\centering
\caption{Log of observations (continued)}
\begin{tabular}{lcccccc}
&&&&&&\\
\hline 
Name & Date of observation  & OBSID & Filter & MJD  & MJD  & Net exposure  \\
     &  (dd-mm-yyyy) &     &       & Start & End  &  time (s)\\
\hline
1ES 1959+650   & 05-10-2016 & A02\_199T01\_9000000708 & F154W & 57666.211111  & 57666.358387  & 2989   \\
               &            &                       & N263M & 57666.211060  & 57666.358438  & 3016   \\
               & 04-11-2016 & G06\_086T01\_9000000774 & F154W & 57695.927007  & 57698.424637  & 16236  \\
               &            &                       & N279N & 57695.926935  & 57698.424604  & 16575  \\  
               & 16-11-2016 & A02\_199T01\_9000000800 & F154W & 57708.444207	 & 57708.452211  & 662    \\ 
               &            &                       & N263M & 57708.444207  & 57708.585130  & 2504   \\
               & 25-10-2017 & T01\_200T01\_9000001638 & F154W & 58051.181646  & 58051.875679  & 14906  \\
               &            &                       & N279N & 58051.181597  & 58051.875729  & 15000  \\
PKS 2005$-$489 & 15-10-2019 & A07\_041T01\_9000003234 & F148W & 58771.266239  & 58771.465491  & 2182   \\
               &            &                       & F154W & 58771.467225  & 58771.673344  & 3806   \\
               &            &                       & F169M & 58771.675082  & 58771.881982  & 3510   \\                
CTA 102        & 21-07-2017 & G05\_218T01\_9000001394 & F148W & 57954.350649  & 57955.900967  & 31074  \\
               &            &                       & N219M & 57954.350598  & 57955.901018  & 27872  \\ 
               & 07-08-2019 & A05\_160T01\_9000003078 & F148W & 58702.177816  & 58702.383449  & 1924   \\
               &            &                       & F154W & 58702.385187  & 58702.395169  & 811    \\
               &            &                       & F172M & 58702.468388  & 58702.603028  & 3509   \\
3C 454.3       & 20-10-2016 & A02\_149T01\_9000000740 & F154W & 57680.625139  & 57681.323406  & 9502   \\
               &            &                       & F172M & 57681.325185  & 57681.752560  & 8068   \\
               &            &                       & N219M & 57680.625089  & 57681.323456  & 7683   \\
               &            &                       & N279N & 57681.325135  & 57681.752560  & 8111   \\
               & 27-06-2023 & T05\_124T01\_9000005712 & F154W & 60121.662641  & 60121.669951  & 595   \\
               &            &                       & F169M & 60121.671689  & 60121.809738  & 2855  \\
               &            &                       & F172M & 60121.811476  & 60122.548271  & 6035  \\   
1ES 2322$-$409 & 03-07-2020 & A09\_147T01\_9000003754 & F148W & 59033.450076  & 59033.664081  & 4296   \\
1ES 2344+514   & 22-10-2017 & A04\_049T01\_9000001632 & F172M & 58048.015081 & 58048.285228  &  1876   \\
               &            &                       & N245M & 58047.952626 & 58048.285277  &  1797   \\  
               & 29-07-2021 & A10\_056T01\_9000004598 & F148W & 59423.293892 & 59423.775796  &  11135  \\
               &            &                       & F154W & 59423.777535 & 59424.582497  &  10642  \\
               &            &                       & F169M & 59424.584238 & 59424.927427  &  5567   \\
               & 06-08-2021 & A10\_056T01\_9000004626 & F148W & 59431.677512 & 59432.358599  &  10791   \\
               &            &                       & F154W & 59432.360333 & 59432.764400  &  8178   \\
               &            &                       & F169M & 59432.909076 & 59433.052119  &  2087   \\    
H 2356$-$309   & 17-10-2018 & A05\_174T01\_9000002440 & F154W & 58407.368693 & 58408.794050  & 18452   \\
               & 18-10-2023 & A05\_174T01\_9000005880 & F154W & 60234.470743 & 60235.833435  & 24817   \\
\hline
\end{tabular}
\label{Table-2}
\end{table*}

\begin{table*}
\centering
\caption{Flux variability analysis results for the sources that show variability. Here, N is the number of orbits, $\sigma_m$ is the intrinsic amplitude of flux variability in magnitude, $F_{var}$ is the fractional root mean square variability amplitude and Err($F_{var}$) is the uncertainty in $F_{var}$}.
\begin{tabular}{lclrccc}
&&&&\\
\hline
Name & Date & Filter & N  & $\sigma_m$  &  $F_{var}$    & Err($F_{var}$)  \\
& (dd-mm-yyyy)  &    &          &  (mag) &  (counts/s)    &   (counts/s)       \\
\hline
1ES 0120+340    & 04-12-2018  & F154W  & 44  & 0.022 & 0.022 & 0.031  \\
1ES 0229+200    & 09-08-2021  & F154W  & 14 & 0.036 & - & - \\
1H 0323+342     & 04-02-2022  & F148W  & 10  & 0.010 & 0.013 & 0.021  \\
1ES 0347$-$121  & 21-08-2020  & F148W  & 04  & 0.072 & 0.063 & 0.031 \\
PKS 0447$-$439  & 15-12-2017  & F154W  & 09  & 0.022 & 0.023 & 0.008 \\
                &             & N245M  & 09  & 0.010 & 0.007 & 0.005 \\
1H 0658+595     & 19-11-2016  & F172M  & 04  & 0.102 & 0.087 & 0.067  \\
OJ 287          & 25-10-2020  & F154W  & 85  & 0.054 & 0.053 & 0.005 \\
                & 20-11-2021  & F154W  & 11  & 0.027 & 0.027 & 0.011 \\
                & 18-04-2023  & F154W  & 16  & 0.055 & 0.052 & 0.012 \\
Mrk 501         & 25-03-2022  & F154W  & 30  & 0.015 & 0.017 & 0.003 \\  
                & 23-10-2022  & F148W  & 54  & 0.027 & 0.026 & 0.003 \\
Mrk 180         & 25-10-2018  & F148W  & 12  & 0.023 & 0.022 & 0.012 \\
                &             & F154W  & 05  & 0.023 & 0.022 & 0.020 \\
Ton 599         & 24-04-2020  & F148W  & 05  & 0.059 & 0.052 & 0.029 \\
                &             & F169M  & 11  & 0.168 & 0.145 & 0.025  \\
                &             & F172M  & 06  & 0.199 & 0.172 & 0.031 \\  
4C +21.35       & 25-04-2018  & F148W  & 02  & 0.065 & 0.045 & 0.022   \\
OQ 334          & 12-01-2020  & F148W  & 05  & 0.121 & 0.116 & 0.008 \\ 
                &             & F154W  & 07  & 0.168 & 0.143 & 0.011 \\
                &             & F169M  & 05  & 0.099 & 0.097 & 0.012 \\ 
S3 1741+19      & 23-08-2019  & F154W  & 08  & 0.029 & 0.009 & 0.228 \\
                &             & F172M  & 08  & 0.167 & 0.171 & 0.048  \\  
1ES 1959+650    & 04-11-2016  & F154W  & 18  & 0.040 & 0.039 & 0.004  \\
                &             & N279N  & 18  & 0.044 & 0.041 & 0.006  \\
                & 25-10-2017  & F154W  & 11  & 0.014 & 0.016 & 0.005  \\
                &             & N279N  & 11  & 0.046 & 0.041 & 0.005 \\ 
PKS 2005$-$489  & 15-10-2019  & F169M  & 04  & 0.013 & 0.013 & 0.018  \\
CTA 102         & 21-07-2017  & F148W  & 21  & 0.062 & 0.058 & 0.010  \\
                &             & N219M  & 21  & 0.035 & 0.034 & 0.017  \\
                & 07-08-2019  & F172M  & 02  & 0.247 & 0.220 & 0.052 \\
3C 454.3        & 27-06-2023  & F172M  & 08  & 0.078 & 0.062 & 0.072   \\ 
1ES 2344+514    & 29-07-2021  & F169M  & 05  & 0.062 & 0.054 & 0.047    \\ 
                & 06-08-2021  & F148W  & 11  & 0.053 & 0.041 & 0.042 \\               
\hline
\end{tabular}
\label{Table-3}
\end{table*}

\section{Flux and Spectral Variability Analysis}
\label{sec:Analysis}

\subsection{Flux Variability}
\label{sec:flux varialbility analysis}
To characterize the flux variability of the sources, we computed the intrinsic variability amplitude, defined as the variance of the observed lightcurve after the removal of the measurement uncertainties ($\sigma_m$, in magnitudes; see Eq. \ref{Eq:Amplitude}). This was done following the methodology described by \cite{ai2010dependence}. 

\begin{equation}
{\sigma_m = \sqrt {(\Sigma^2 - \epsilon^2)}}
\label{Eq:Amplitude}
\end{equation}

where, $\Sigma^2$, the variance of observed magnitudes, is given by 

\begin{equation}
\Sigma = \sqrt{\frac{1}{N-1} \sum_{i=1}^{N} (m_i - \langle m \rangle)^2}
\end{equation}

and $\epsilon^2$, the mean error square due to measurement errors, is given by 

\begin{equation}
\epsilon^2 = \frac{1}{N}\sum_{i=1}^{N} \epsilon_i^2
\end{equation}

Here, N and $\langle m \rangle$ are the number of orbits and the weighted mean of $m_i$ measurements with the uncertainties $\epsilon_i$, respectively. We considered a source to be variable only if  $\Sigma^2$ $>$ $\epsilon^2$, otherwise the intrinsic amplitude of flux variability $\sigma_m$ = 0, and the source is non-variable.   
\\[6pt]
We also calculated the fractional root mean square variability amplitude ($F_{var}$) to characterise flux variability. $F_{var}$ is defined as \citep{vaughan2003characterizing}

\begin{equation}
    F_{var} = \sqrt{\frac{S^2 - \overline{\sigma_{err}^2}}{\bar{x}^2}}
\end{equation}

Here, $S^{2}$ is the sample variance of the light curve in counts/s, $\overline{x}$ is the mean of the light curve and $\overline{\sigma_{err}^2}$ is the mean square error. They are defined as:

\begin{equation}
    S^2 = \frac{1}{N-1} \sum_{i=1}^N (x_{i}-\bar{x})^2
\end{equation}

\begin{equation}
    \overline{\sigma_{err}^2} = \frac{1}{N} \sum_{i=1}^N \sigma_{err,i}^2
\end{equation}

where $x_{i}$ is the flux of individual data point in counts/s and $\sigma_{err,i}$ is the measured uncertainty on each data point. The error in $F_{var}$ is defined as \citep{vaughan2003characterizing}

\begin{equation}
    Err (F_{var}) = \sqrt{\left(\sqrt{\frac{1}{2N}}\frac{\overline{\sigma_{err}^2}}{\bar{x}^2F_{var}}\right) ^2  + \left(\sqrt{\frac{\overline{\sigma_{err}^2}}{N}} \frac{1}{\bar{x}}\right)^2}
\end{equation}

The flux variability results are presented in Table \ref{Table-3} for the sources that exhibit variability according to Eq. (1).

\subsection{Spectral Variability}
\label{sec: spectral variability analysis}

\begin{figure*}
\hbox{
\hspace{2 cm}
\includegraphics[scale=0.50]{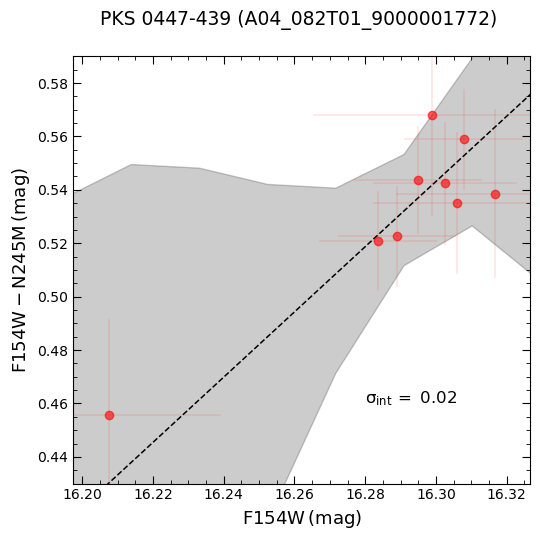}
\hspace{0.5 cm}
\includegraphics[scale=0.50]{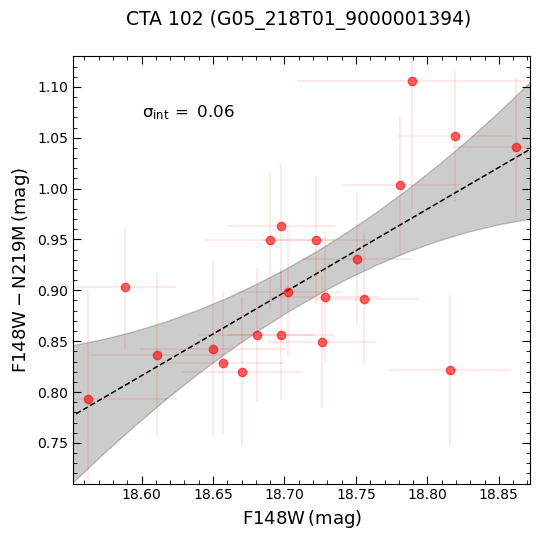}
}
\vspace{0.5 cm}
\hbox{
\hspace{2 cm}
\includegraphics[scale=0.50]{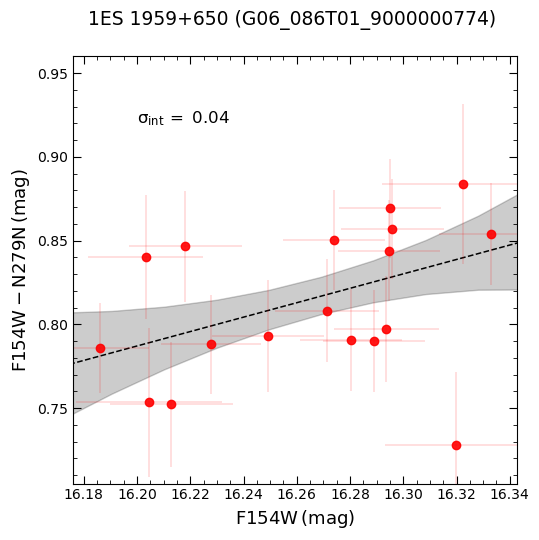}
\hspace{0.5 cm}
\includegraphics[scale=0.50]{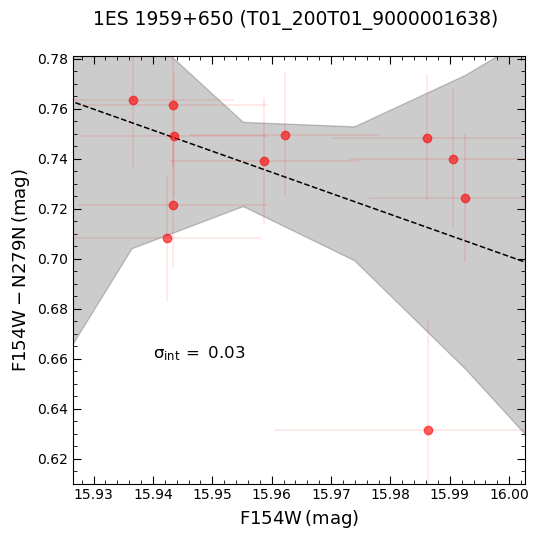}
}
\caption{The CMDs of the sources analyzed for spectral variability. The black dashed line and the grey shaded region represent the best fit and uncertainty range on the fit from Bayesian analysis using LINMIX$\_$ERR method. The intrinsic scatter ($\sigma_{int}$) value is provided in each plot and the name of the source along with the OBSID is provided on the top of each plot (also refer Table \ref{Table-4} and \ref{Table-5}).}
\label{figure-6}
\end{figure*}

To investigate the spectral variability characteristics of the sources analysed in this work, we constructed colour magnitude diagrams (CMDs).  In the optical band, the two  subclasses of blazars are known to exhibit distinct trends in their CMDs: BL Lacs generally show a bluer-when-brighter (BWB) behaviour, whereas FSRQs tend to display a redder-when-brighter (RWB) trend \citep{negi2022optical, li2024optical}. In this study, we analyzed color variations for sources that have more than four simultaneous FUV and NUV photometric measurements on hour timescales. We fitted the CMDs of these sources using both unweighted and weighted linear least-squares fit methods. For the weighted linear least-squares fitting, which accounts for uncertainties in both the color index and magnitude, as well as their correlated errors, we employed two approaches, the Bivariate Correlated Errors and Intrinsic Scatter (BCES) method \citep{akritas1996linear}, and LINMIX$\_$ERR, a Bayesian linear regression technique \citep{kelly2007some}. The fitting was performed using the following form : colour = slope $\times$ (m - $m_{0}$) + intercept, where $m_{0}$ is the median of the magnitude values.The spectral fit plots are shown in Fig. \ref{figure-6} and the results of the fit are presented in Table \ref{Table-4} and Table \ref{Table-5}. A positive slope in the CMD indicates a BWB trend, while a negative slope corresponds to a RWB trend. In addition, we considered a source to exhibit significant spectral variability if the absolute value of the linear correlation coefficient satisfies $|R|>$ 0.5 and the probability of no correlation $P<$ 0.05.

\begin{table*}
\centering
\caption{Spectral variability results determined using the unweighted linear least-squares fitting method. Here, R is the linear correlation coefficient and P is the probability for no correlation.}
\begin{tabular}{l@{\hspace{0.3cm}}c@{\hspace{0.2cm}}c@{\hspace{0.2cm}}c@{\hspace{0.25cm}}c@{\hspace{0.25cm}}l@{\hspace{0.25cm}}c@{\hspace{0.25cm}}c@{\hspace{0.25cm}}}
&&&&&&&\\
\hline
Name of source  &  OBSID & N  & color/mag & Slope & Intercept & R & P  \\
      &        &&           & & & & \\
\hline
PKS 0447$-$439 & A04\_082T01\_9000001772 & 09                   & (F154W$-$N245M)/F154W & 0.91$\pm$0.16  & 0.54$\pm$0.01 &  0.91 & 0.00     \\
1ES 1959+650   & G06\_086T01\_9000000774 & 18 & (F154W$-$N279N)/F154W & 0.34$\pm$0.23 & 0.82$\pm$0.01 & 0.35 & 0.16  \\
               & T01\_200T01\_9000001638 & 11                   & (F154W$-$N279N)/F154W & $-$0.61$\pm$0.51  & 0.73$\pm$ 0.01 & $-$0.37 &  0.26     \\
CTA 102        & G05\_218T01\_9000001394 & 21 & (F148W$-$N219M)/F148W & 0.73$\pm$0.20  & 0.90$\pm$0.02 &  0.65 & 0.00 \\
\hline
\end{tabular}
\label{Table-4}
\end{table*}

\begin{table*}
\centering
\caption{Spectral variability results determined using the BCES and Bayesian methods. Here, Columns 5 and 6 are the slope and intercept values along with their associated errors calculated based on BCES method, Column 7 and 8 are the slope and intercept values along with their associated errors calculated based on Bayesian method.}
\begin{tabular}
{l@{\hspace{0.1cm}}c@{\hspace{0.1cm}}c@{\hspace{0.05cm}}c@{\hspace{0.1cm}}c@{\hspace{0.1cm}}c@{\hspace{0.1cm}}c@{\hspace{0.1cm}}c@{\hspace{0.1cm}}} 
&&&&&&&\\
\hline
Name  &  OBSID & N & color/mag & \multicolumn{2}{c}{BCES} & \multicolumn{2}{c}{Bayesian} \\
      &        &               &           &   Slope    &  Intercept   &   Slope    &  Intercept       \\ 
\hline
PKS 0447$-$439 & A04\_082T01\_9000001772 & 09                   & (F154W$-$N245M)/F154W & 0.91$\pm$0.09 & 0.54$\pm$0.01 & 1.08$\pm$4.00 & 0.54$\pm$0.04  \\
1ES 1959+650   & G06\_086T01\_9000000774 & 18 & (F154W$-$N279N)/F154W & 0.55$\pm$0.24 &  0.82$\pm$0.01 & 0.44$\pm$0.32 & 0.82$\pm$0.01  \\
               & T01\_200T01\_9000001638 & 11                & (F154W$-$N279N)/F154W & $-$1.73$\pm$1.41 & 0.74$\pm$0.02 & $-$0.54$\pm$8.01 & 0.73$\pm$0.03 \\
CTA 102        & G05\_218T01\_9000001394 & 21 & (F148W$-$N219M)/F148W & 0.78$\pm$0.19 & 0.90$\pm$0.01 & 0.80$\pm$0.40 & 0.90$\pm$0.02 \\ 
\hline
\end{tabular}
\label{Table-5}
\end{table*}

\begin{figure*}
\centering
\hbox{
\hspace{0.5cm}
\includegraphics[scale=0.62]{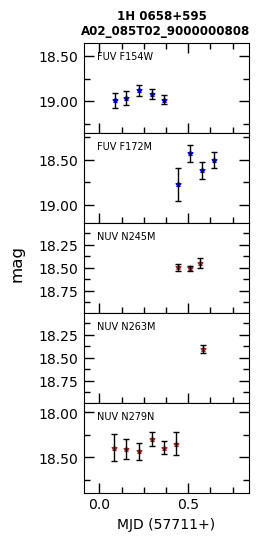}
\includegraphics[scale=0.62]{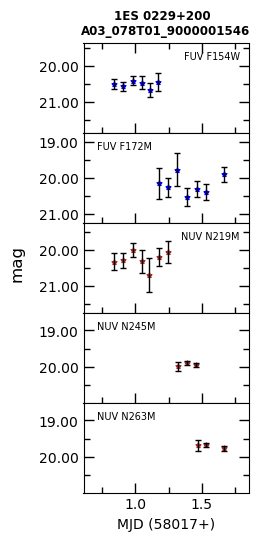}
\includegraphics[scale=0.62]{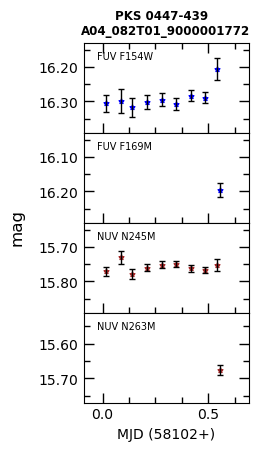}
\includegraphics[scale=0.62]{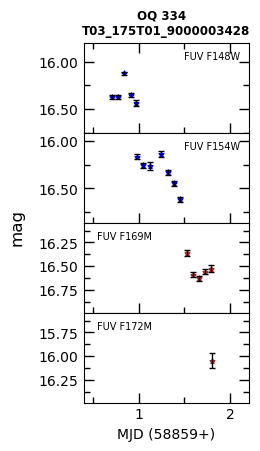}
}
\vspace{0.15cm}
\hbox{
\hspace{0.5cm}
\includegraphics[scale=0.62]{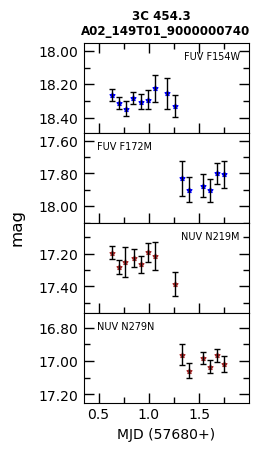}
\includegraphics[scale=0.62]{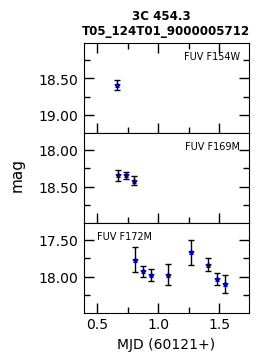}
\includegraphics[scale=0.62]{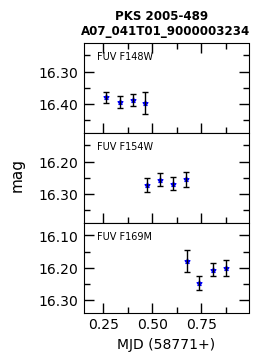}
\includegraphics[scale=0.62]{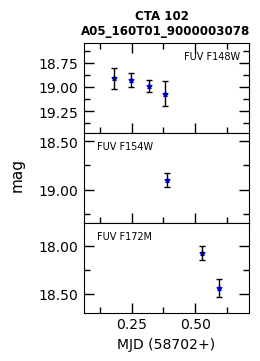}

}
\vspace{0.15cm}
\hbox{
\hspace{0.5cm}
\includegraphics[scale=0.62]{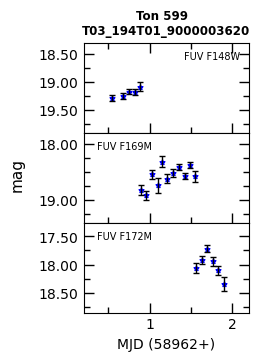}
\includegraphics[scale=0.62]{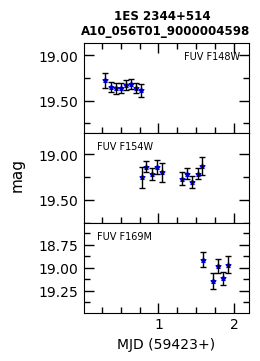}
\includegraphics[scale=0.62]{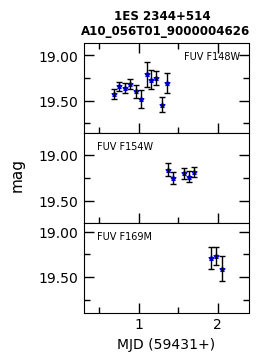}
\includegraphics[scale=0.62]{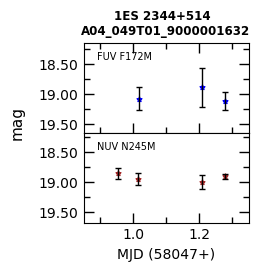}
}
\caption{FUV (blue stars) and NUV (brown stars) lightcurves of sources. The source name along with the observation ID and filter names are provided in the respective plots.}
\label{figure-3}
\end{figure*}

\begin{figure*}
\centering
\hbox{
\hspace{0.5cm}
\includegraphics[scale=0.62]{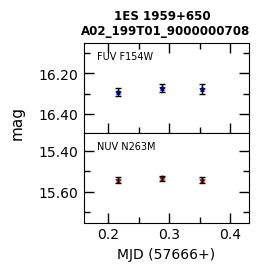}
\includegraphics[scale=0.62]{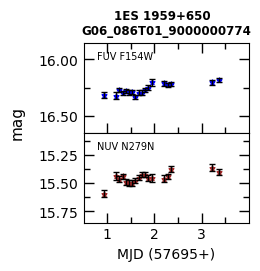}
\includegraphics[scale=0.62]{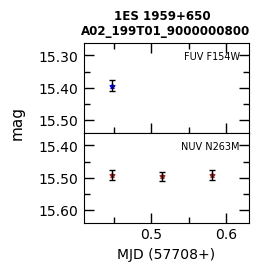}
\includegraphics[scale=0.62]{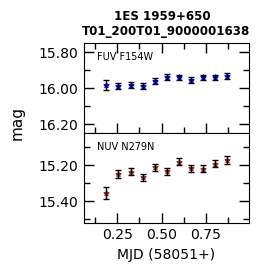}
}
\vspace{0.5cm}
\hbox{
\hspace{0.5cm}
\includegraphics[scale=0.62]{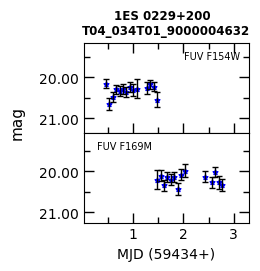}
\includegraphics[scale=0.62]{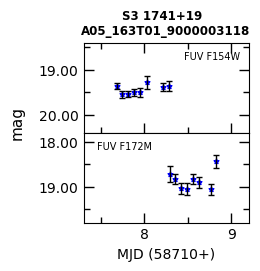}
\includegraphics[scale=0.62]{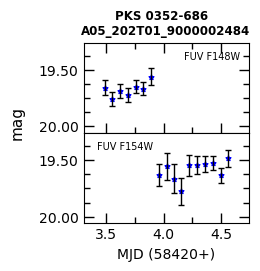}
\includegraphics[scale=0.62]{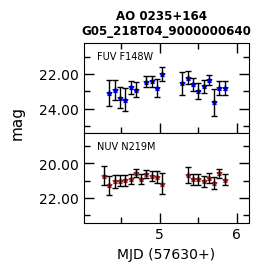}
}
\vspace{0.5cm}
\hbox{
\hspace{0.5cm}
\includegraphics[scale=0.62]{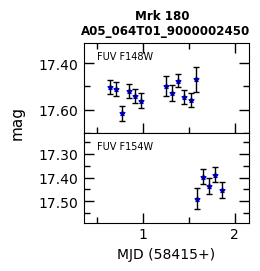}
\includegraphics[scale=0.62]{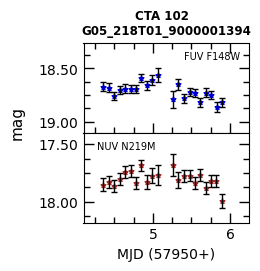}
\includegraphics[scale=0.62]{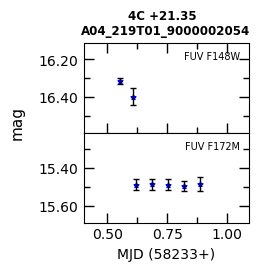}
\includegraphics[scale=0.62]{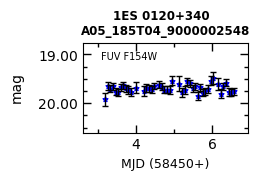}
}
\vspace{0.5cm}
\hbox{
\hspace{0.5cm}
\includegraphics[scale=0.62]{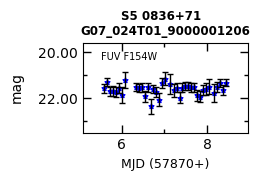}
\includegraphics[scale=0.62]{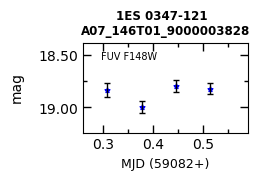}
\includegraphics[scale=0.62]{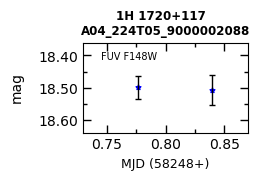}
\includegraphics[scale=0.62]{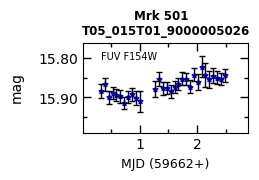}
}
\vspace{0.5cm}
\hbox{
\hspace{0.5cm}
\includegraphics[scale=0.62]{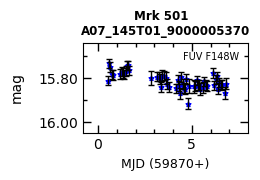}
\includegraphics[scale=0.62]{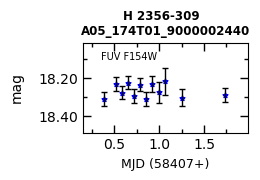}
\includegraphics[scale=0.62]{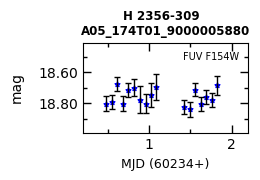}
\includegraphics[scale=0.62]{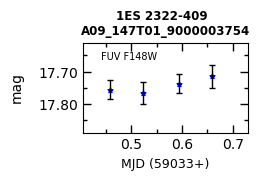}
}

\caption{FUV (blue stars) and NUV (brown stars) lightcurves of sources. The source name along with the observation ID and filter names are provided in the respective plots.}
\label{figure-4}
\end{figure*}

\begin{figure*}
\centering
\hbox{
\hspace{0.5cm}
\includegraphics[scale=0.62]{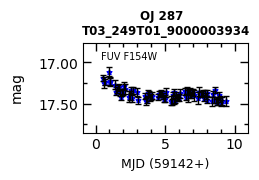}
\includegraphics[scale=0.62]{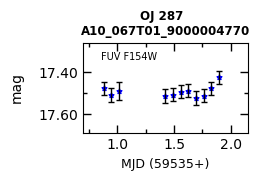}
\includegraphics[scale=0.62]{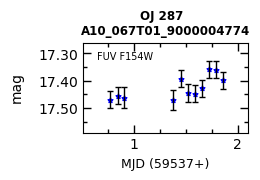}
\includegraphics[scale=0.62]{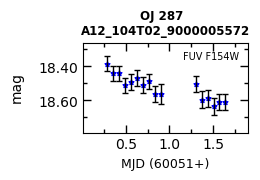}
}
\vspace{0.5cm}
\hbox{
\hspace{0.5cm}
\includegraphics[scale=0.62]{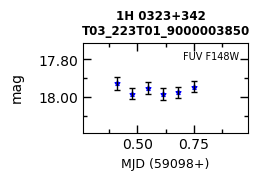}
\includegraphics[scale=0.62]{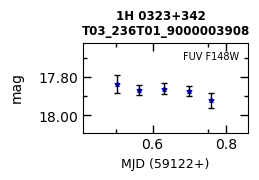}
\includegraphics[scale=0.62]{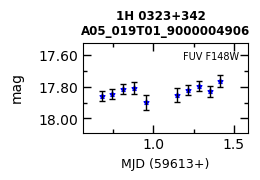}
}
\caption{FUV (blue stars) lightcurves of sources. The  source name along with the observation ID and filter names are provided in the respective plots.}
\label{figure-5}
\end{figure*}

\section{Notes on individual sources}
\label{sec:Individual source description}

\subsection{1ES 0120+340}
The source 1ES 0120+340, located at a redshift of $z$ = 0.272 \citep{costamante2001extreme}, is classified as a HSP  BL Lac object \citep{goswami2024variety,ackermann2015third}. It has previously been investigated for high-energy $\gamma$-ray emission by \cite{de2003search}. In the present work, we report UV variability on hour timescales for the first time for this source. The source was observed during a single epoch using the FUV F154W filter and it exhibited statistically significant UV flux variability, confirming its intrinsically variable nature. 

\subsection{1ES 0229+200}
1ES 0229+200 is a HSP BL Lac object with a TeV spectrum extending up to 10 TeV \citep{aharonian2007new} and located at a redshift of $z$ = 0.139 \citep{ackermann2015third}. Modest optical variability of $\sim$0.2 mag was reported by \cite{wierzcholska2015longterm} based on monitoring observations during 2007 - 2012. It was found to be variable in X-rays from NuSTAR observations \citep{pandey2017x}. It was found to show rapid intra-night optical variability \citep{bachev2012nature}, and a rapid optical flare lasting for about 6 hours was reported by \cite{kishore2024rapid}. The UV flux variability in this source was first investigated by \cite{reshma2024ultraviolet}. In this work, we re-investigated its UV flux variability using observations from two additional epochs. The source was found to exhibit variability only in the F154W filter during the 2021 epoch. 
\subsection{AO 0235+164}
AO 0235+164 is a LSP BL Lac object located at a redshift of  $z$ = 0.94 \citep{ackermann2015third}. \cite{raiteri2001optical} systematically investigated the source in the radio and optical bands and reported quasi-periodic oscillations with the possible indication of the source to host a binary black hole system \citep{romero2003binary}. Although it is typically classified as a BL Lac object, the source also exhibits several characteristics commonly associated with FSRQs \citep{wang2020comprehensive}. AO 0235+164 is well known  as an optically violent variable source \citep{roy2023study} and its long-term, multi-wavelength variability, has been studied extensively by \cite{vlasyuk2024multiwavelength}. It has been extensively studied for variability across wavelengths and has been found to show flux variations on a range of timescales \citep{2000A&A...360L..47R,raiteri2001optical,ackermann2012multi}. This source was observed simultaneously in the FUV and NUV bands on 4 September 2016. During this epoch the source did not show variability in either the FUV or NUV bands.

\subsection{1H 0323+342}
1H 0323+342 was first discovered in the \textit{HEAO-1} survey by \cite{wood1984heao}. It is a radio-loud NLSy1 galaxy, notable for being one of the nearest ($z$ = 0.063) and optically brightest members of this class as well as one of the few radio-loud NLSy1 galaxies detected at $\gamma$-ray energies \citep{turner2022optical}. Radio-loud NLSy1 galaxies are known to share several observational properties with blazars \citep{abdo2009radio}. The source hosts a central black hole with an estimated  mass of $\sim$$10^7$ $M_\odot$ \citep{zhou2007narrow}. The source was found to be variable on intra-night time scales in the optical band \citep{paliya2013intranight,paliya2014peculiar,ojha2019intra}. \cite{feigelson1986h} reported rapid variability in the X-ray band with a halving timescale of 30 seconds. From long term variability analysis using observations from \textit{Swift}, \cite{d2020gamma} found correlated variations in the optical, UV and X-ray bands favouring the contribution of the jet to the observed emission in those bands. This source was observed with UVIT over three epochs in the FUV band using the F148W filter. Of these three epochs, significant UV  flux variabilities were detected during the observations carried out on 4 February 2022, representing the first report of UV variability on this source. 

\subsection{1ES 0347$-$121}
1ES 0347$-$121 located at a redshift of $z$ = 0.188 \citep{woo2005black} is a member of the Einstein Slew Survey \citep{elvis1992einstein} and was classified as a BL Lac object by  \cite{schachter1993ten}. The  source is also known to be an extreme TeV blazar \citep{goswami2024variety}. Its X-ray flux and spectral variability have previously been studied by \cite{pandey2018x} and was found to be non-variable in X-rays. The source was found to be variable in the optical from long term monitoring observations \citep{wierzcholska2015longterm}. In this work, we report, for the first time, UV variability in this source on timescales of hours. The source  was observed on 21 August 2022 in the FUV band and was found to be variable.

\subsection{PKS 0352$-$686}
PKS 0352$-$686 is a BL Lac type blazar located at a redshift of $z = 0.087$ \citep{masetti2006unveiling}. From observations carried out with the Transiting Exoplanet Survey Satellite, \cite{tripathi2026probable} reported detection of quasi-periodic oscillation in this source. UV observations of this  source carried out in 2018 using two FUV filters revealed no significant variability during that epoch. 

\subsection{PKS 0447$-$439}
PKS 0447$-$439 is a BL Lac type blazar \citep{muriel2015bl} located at a redshift of $z$ = 0.205 \citep{perlman1998deep}. The source was  originally discovered in the radio band by \cite{large1981molonglo} and has since been extensively observed across multiple wavelengths, including radio \citep{gregory1994parkes}, X-rays \citep{white1994wgacat}, UV \citep{lampton1997all}, and high-energy $\gamma$-rays 
by \textit{Fermi}-LAT \citep{abdo2009bright}. It is believed to be powered by a supermassive black hole with a mass of approximately $6 \times 10^8$ $M_\odot$ \citep{ghisellini2010general}. In this work, we report for the first time the detection of UV variability in PKS 0447$-$439 on hour timescales. It was observed on 15 December 2017, during which  variability was detected in both the F154W and N245M filters.

\subsection{1H 0658+595}
1H 0658+595 ($z$ = 0.125) is classified as a HSP blazar and is known to be an extreme TeV emitter \citep{furniss2013blazar}. A long-term investigation of its X-ray and $\gamma$-ray variability by \cite{goswami2024variety} revealed significant flux and spectral variations, with X-ray variability observed on timescales ranging from days to weeks. In this work, we searched for UV flux variability on hour timescales, however, the source was found to be variable only in the F172M observation.

\subsection{S5 0836+71}
S5 0836+71, is a FSRQ \citep{malizia2000hard} located at a high redshift of $z$ = 2.218 \citep{ackermann2015third}. It is classified as a low-polarization quasar in the optical band \citep{impey1990optical}, and is known to host a supermassive black hole with a mass of $3 \times 10^9$ $M_\odot$ \citep{ghisellini2010chasing} and $5 \times10^9$ $M_\odot$ \citep{tagliaferri2015nustar}. The source exhibits a prominent big blue bump in its SED indicative of strong thermal emission from the accretion disk \citep{raiteri2014infrared, raiteri2019beamed}. \cite{orienti2020radio} investigated the polarisation and multi-wavelength variability of S5 0836+71 from radio to $\gamma$-rays using VLBA, \textit{Swift} and \textit{Fermi-LAT} observations and reported that the optical-UV emission was largely dominated by the accretion disk exhibiting lesser variability. In this study, we examined the source's UV flux variability on hour timescales using UVIT observations obtained on 4 May 2017. No variability was detected during this epoch.

\subsection{OJ 287}
OJ 287 is a LSP BL Lac type blazar at a redshift of $z$ = 0.306. It is a well-known blazar to host a binary supermassive black hole \citep{sillanpaa1988oj} detected with a high optical activity at a periodic interval of 12 years \citep{prince2021comprehensive}. OJ 287 is detected in TeV \citep{2017arXiv170802160O} and has been extensively studied for variability across wavelengths. It has been found to show X-ray flux and spectral variations \citep{2024MNRAS.532.3285Z}. It is variable in the optical with \cite{2024ApJ...960...11K} reporting a shortest variability time scale of 0.38 days. It has also been found to be variable in infra-red \citep{2022ApJS..260...39G}, as well as in the optical and $\gamma$-ray bands \citep{2021MNRAS.504.1772R}. In addition to the detailed UV flux variability analysis reported by \cite{reshma2024ultraviolet}, we analyzed four new epochs of observations carried out with UVIT in this work. We found the source to show UV flux variability on three epochs of observations.  

\subsection{Mrk 180} 
Mrk 180 is a HSP blazar located at a redshift of $z$ = 0.045. It was first identified as a BL Lac type object in 1976 \citep{mondal2023exploring}. It was initially detected in X-rays by \cite{hutter1981optical}, and its very high energy TeV $\gamma$-ray emission was subsequently discovered by \cite{albert2006discovery}. Its optical long term variability has been analysed by \cite{nilsson2018long}. Mrk 180 was observed by UVIT on 25 October 2018, using two FUV filters, F148W and F154W. We found the source to be variable in both the filters.

\subsection{Ton 599}
Ton 599 is a LSP FSRQ located at a redshift of $z$ = 0.725 \citep{rajput2024investigation} and it is a TeV $\gamma$-ray source \citep{mirzoyan2017detection}. The source is classified as both an optically violent variable and a highly polarized quasar \citep{patel2018temporal}. It exhibits variability across a broad energy range, from radio to $\gamma$-rays, and has been the subject of extensive flux variability studies for more than three decades \citep{manzoor2024broad}. A multiwavelength variability study conducted by \cite{prince2019multi} during the 2017 high-activity state reported strong $\gamma$-ray variability and a comparatively less variability in the X-ray band. It was found to show strong optical flux variations \citep{vince2025multiband}. It was found to show correlated optical and GeV flux variations at certain epochs, while such correlated variations are not seen at all times \citep{rajput2024investigation}. In this work, we analyzed UVIT observations of Ton 599 obtained  on 24 April 2022, using three different FUV filters. UV flux variability was detected in all the three filters. 

\subsection{4C +21.35}
4C +21.35 is FSRQ type blazar located at a redshift of $z$ = 0.435 \citep{ackermann2015third}. It is a well known $\gamma$-ray emitting source \citep{hartman1999third}, and hosts a central black hole with an estimated mass of $\sim$6 $\times$ $10^8$ $M_\odot$ \citep{farina2012optical}. It has been analysed for X-ray flux variation by \cite{2016MNRAS.458...56W}. Outburst in the GeV band was reported by \cite{2010ATel.2584....1D}, and in the high energy $\gamma$-ray band flux variations with a doubling time scale of about 10 minutes was reported by \cite{aleksey2011magic}. The broadband SED of the source has been studied in detail by \cite{ackermann2014multifrequency}. The source  was observed with UVIT on 25 April 2018 using two FUV filters, F148W and F172M. UV variability was detected in the F148W filter.  

\subsection{OQ 334}
OQ 334, is a FSRQ located at a redshift of $z$ = 0.682. It is a very high energy $\gamma$-ray source detected by the \textit{Fermi}-LAT \citep{dammando2021multiwavelength}. Multi-wavelength studies, covering radio to very high energy $\gamma$-rays have been carried out on this source \citep{dammando2021multiwavelength, ren2024oscillation}. Until 2018, OQ 334 remained largely in a quiescent state \citep{abdollahi2020fermi}, after which it underwent a series of flaring episodes accompanied by significant changes in its broad emission lines. These spectral transformations led \cite{mishra2021changing} to classify the source as a changing-look type blazar, owing to its transition between FSRQ and BL Lac characteristics. It was also found to show repeated flaring activity in the GeV $\gamma$-ray band \citep{ciprini2020fermi}. In this work, we investigated hour UV variability in OQ 334 for the first time. UVIT observations carried out on 12 January 2020, using four different FUV filters revealed significant variability in three of them, indicating a variable nature in the UV band.

\subsection{Mrk 501}
Mrk 501 is a HSP BL Lac object at a redshift of $z$ = 0.034. It has been studied for optical/UV and X-ray variability on timescales ranging from months to years by \cite{kapanadze2023long} and \cite{tantry2024probing}. Long term variability study in the X-ray and $\gamma$-ray band was studied by \cite{gliozzi2006long}. Rapid optical variability on minutes timescale was reported by \cite{zeng2019minute}. It has been extensively studied for UV variability on hour timescales by \cite{reshma2024ultraviolet}. In 2022, UVIT observed the source during two separate epochs using different FUV filters, and UV  variability was detected in both observations.
 
\subsection{1H 1720+117}
1H 1720+117, located at redshift $z$ = 0.018, is classified as a BL Lac type blazar. \citet{nieppola2006spectral} categorized it as an ISP blazar, with a synchrotron peak frequency at $\nu_{s} = 6.3\times10^{15}$ Hz. Optical monitoring by \citet{kalita2021optical} reported no intra-day variability in its lightcurve. Interestingly, despite its BL Lac classification, the source exhibited a pronounced RWB trend in the optical band. During UVIT observations carried out on 10 May 2018 using the FUV filter F148W, the source was found to be non-variable. 

\subsection{S3 1741+19}
S3 1741+19, identified in the \textit{Einstein} Slew Survey \citep{perlman1996einstein}, is a HSP blazar located at a redshift of $z$ = 0.084. It is hosted by one of the largest and most luminous elliptical host galaxies among BL Lac objects. The host galaxy appears to have been tidally distorted due to interactions with two companion galaxies, forming a dynamically interacting triplet system \citep{abeysekara2016veritas, ahnen2017magic}. The source was detected in the high-energy MeV $\gamma$-ray regime by \textit{Fermi}-LAT \citep{abdo2010fermi}, while its very-high-energy $\gamma$-ray emission was first reported by \cite{berger2011overview}. Optical observations by \cite{heidt19991es} provided an early detailed study of the source, and \cite{abeysekara2016veritas} reported variability in both the optical and UV bands. In contrast, \cite{goswami2024variety} found no significant variability in the optical and $\gamma$-ray bands, but reported clear  variability at X-ray energies. On 23 August 2019, UVIT observed S3 1741+19 using two FUV filters, F154W and F172M. We detected UV variations on hour timescales in both the filters.

\subsection{1ES 1959+650}
The blazar 1ES 1959+650 was first detected in the X-ray band by the \textit{Einstein} Slew Survey \citep{elvis1992einstein} and subsequently classified as a BL Lac object by \cite{schachter1993ten}. Located at a redshift of $z$ = 0.047, it is categorized as a HSP blazar \citep{krawczynski2004multiwavelength}. The source is a known $\gamma$-ray emitter, with its first detection in the TeV regime reported by \cite{nishiyama1999detection}, followed by detailed studies of a TeV outburst observed with  VERITAS \citep{holder2002detection}. A multi-wavelength variability analysis was presented by \cite{aliu2014investigating}. In this work, we investigated the UV variability on hour timescales using four epochs of observations with UVIT. We observed flux variations in both FUV and NUV during two epochs, namely 4 November 2016 and 25 October 2017. However, we found no statistically significant spectral variations on both the epochs.

\subsection{PKS 2005$-$489}
PKS 2005$-$489 was first identified as a radio source in the Parkes 2.7 GHz survey by \cite{wall1975parkes} and later classified as a BL Lac object at a redshift of $z$ = 0.071 \citep{ wall1986pks, falomo1987redshift}. It is recognized as a HSP blazar \citep{acero2010pks}. Spectral variability in the X-ray band was studied by \cite{perlman1999x}, while \cite{dominici2004long} investigated its long-term optical variability. The H.E.S.S. Cherenkov telescope array reported the first detection of very-high-energy $\gamma$-ray emission from the source \citep{aharonian2005discovery}. During a UVIT observation on 15 October 2019, the source was monitored using three FUV filters. Variability was detected only in the F169M filter.

\subsection{CTA 102}
CTA 102, located at a redshift of $z$ = 1.037, is classified as a FSRQ \citep{schmidt1965large} and hosts a central black hole with a  mass 
8.5 $\times$ $10^8$ M$_\odot$ \citep{zamaninasab2014dynamically}. It is also categorized as a highly polarized quasar, exhibiting optical polarization levels exceeding 3\% \citep{moore1981class}. The source has been studied for long-term flux and spectral variability across multiple wavelengths, including radio, optical–UV, X-rays, and $\gamma$-rays \citep{d2019investigating}. CTA 102 is notable for being the third known FSRQ to display extremely rapid $\gamma$-ray flux variability on timescales of only a few minutes \citep{shukla2018short}. The source was observed with  UVIT during two epochs, on  21 July 2017 and 7 August 2019.  Significant UV flux variability was detected on both the epochs. Additionally, during the epoch with  simultaneous FUV and NUV observations, we detected spectral variations characterized by  a significant BWB trend.

\subsection{3C 454.3}
3C 454.3, located at a redshift of $z$ = 0.859, is a well-known and highly variable FSRQ hosting a central black hole with a  mass of about 1.5 $\times$ $10^9$ M$_\odot$ \citep{sahakyan2021modelling}. The source is known for its dramatic variability across the entire electromagnetic spectrum. Its multi-wavelength behaviour has been extensively investigated over the past decades \citep{raiteri2008new, jorstad2013tight, amaya2020multiwavelength, chen2025study}. We examined hour timescale UV variability using UVIT observations carried out on 20 October 2016 and 27 June 2023. UV  variability was detected  during the 2023 epoch, while no variability was observed in the 2016 data.

\subsection{1ES 2322$-$409}
1ES 2322$-$409 is classified as a BL Lac object based on its featureless optical spectrum \citep{thomas1998identification}. Although its redshift remains uncertain, \cite{jones20096df} reported a value of $z$ = 0.174. Based on its synchrotron peak located at $10^{15.92}$ Hz, the source is characterized as a HSP blazar \citep{ackermann2015third}. It is a bright TeV emitter, with its very-high-energy $\gamma$-ray emission first reported by \cite{abdalla2019vhe}. UVIT observations of 1ES 2322$-$409 were carried out on 3 July 2020. The source was found to be non-variable during this epoch. 

\subsection{1ES 2344+514}
1ES 2344+514 is a BL Lac object located at a redshift of $z$ = 0.044. It is found to be variable in X-rays \citep{devanand2022study}. Optical variability of the source was studied by \cite{gaur2012quasi} and \cite{cai2022long}. \cite{abe2024multi} reported long timescale UV variability using \textit{Swift}-UVOT observations. Its UV variability has been studied in detail by \cite{reshma2024ultraviolet}, who reported hour timescale flux variations in the UV band. In the present work, we further examined the source  for variability using three additional epochs of UVIT observations conducted on 22 October 2017, 29 July 2021 and 8 August 2021. The source was found to be variable on two of the three epochs analyzed here. 

\subsection{H 2356$-$309}
The BL Lac object H 2356$-$309 is hosted by an elliptical galaxy at a redshift of $z$ = 0.165 \citep{falomo1991galaxy}. It is classified as a HSP blazar, originally identified in the optical band by \cite{schwartz1989lecture}. The source was first detected in X-rays by the \textit{UHURU} satellite, followed by observations with HEAO-I \citep{forman1978fourth, wood1984heao}. Its very-high-energy $\gamma$-ray emission has been studied using the H.E.S.S. Cherenkov telescope \citep{aharonian2006discovery}. \cite{wani2020x} investigated its flux and spectral variability in the X-ray regime. The source was observed for UV variability on 17 October 2018 and 18 October 2023.  Variability was not detected in the source on both the epochs of observations.

\section{Results and Discussion} 
\label{sec:discussion}
Of the 24 sources analyzed for flux variability, 18 exhibited significant variability, while 6 showed no detectable variations. Among the 18 variable sources, 15 have observations only in the  FUV band, whereas three have simultaneous observations in both the FUV and NUV bands. In the NUV band, the variability amplitude ranges from 0.010 mag to 0.046 mag, while in the FUV band it spans a wider range, from 0.010 mag to 0.247 mag. For two of the three sources with simultaneous FUV and NUV observations, the variability amplitude is larger in the FUV than in the NUV. The flux variability results for the variable sources are presented in Table \ref{Table-3}. The UV lightcurves of the sources analyzed in this work are shown in Figs. \ref{figure-3}, \ref{figure-4}, and \ref{figure-5}. 
\\[6pt]
The low energy hump in the broadband SED of FSRQs (covering radio, infrared, optical and UV wavelengths), and BL Lacs (extending from infrared through optical and UV to X-rays) is dominated by synchrotron emission from relativistic non-thermal electrons in their jets. However, in FSRQs during their faint states, prominent contribution from accretion disk is evident in the low-energy hump of the SED \citep{paliya2016broadband,paliya2017general}. Similarly, BL Lacs may also exhibit signatures of accretion disk emission during low states \citep{raiteri2008radio}. Consequently, the observed UV emission in both FSQRs (with broad emission lines and radiatively efficient accretion disks)  and BL Lacs (with weak or absent emission lines and radiatively inefficient accretion disks) is likely a combination of jet and accretion disk emission. Since thermal emission from the accretion disk has a bluer spectral shape, while non-thermal synchrotron emission has a redder spectral shape, the observed colour variations  in blazars are governed  by changes in the disk emission, jet emission or both. In the optical band,  FSRQs are generally  known to exhibit a RWB trend \citep{bonning2012smarts,sarkar2019long}, whereas BL Lacs typically show a BWB behaviour \citep{d2002spectral,vagnetti2003spectral,fiorucci2004continuum,meng2018multi,gaur2019optical}. However, sources are also known to exhibit both BWB and RWB trends \citep{wu2011optical,rajput2019temporal}.
\\[6pt]
In our sample, constrained by the availability of simultaneous FUV and NUV observations, UV colour variability could be examined  only for three sources, namely CTA 102 (FSRQ), PKS 0447$-$439 and 1ES 1959+650 (both BL Lacs). For 1ES 1959+650, during the observations on 4 November 2016, we found a weak positive correlation (R =  0.35), however, this correlation is not statistically significant (P =  0.16) and the fitted slope (0.34$\pm$0.23) is consistent with zero. The slopes from BCES and LINMIX\_ERR too are consistent with zero (see Table \ref{Table-5}). During the observations on 24 October 2017, we found a weak negative correlation (R = $-$0.37), which is again statistically insignificant (P = 0.26). Thus, no significant colour variability was detected in  1ES 1959+650. In contrast, in  both PKS 0447$-$439 and CTA 102, we found a statistically significant positive correlation between UV colour and magnitude (with R of 0.91 in PKS 0447$-$439 and 0.65 in CTA 102, and P$<$0.01 in both the sources) indicating  a clear BWB trend. Both sources showed mild UV brightening  during the epochs in which colour variations were detected. The observed CMDs along with the fits are shown in Fig. \ref{figure-6}. The results of the fit are given in Tables \ref{Table-4} and \ref{Table-5}.
\\[6pt]
During the epoch of the UVIT observations of CTA 102, the source was in a relatively bright $\gamma$-ray flux state (within $\pm$3 days of UVIT observations) with \textit{Fermi}-LAT flux values\footnote{https://fermi.gsfc.nasa.gov/ssc/data/access/lat/LightCurveRepository} in the range (1.6$-$1.8) $\times$ 10$^{-6}$ photons cm$^{-2}$ s$^{-1}$. In contrast, PKS 0447$-$439 was not detected in $\gamma$-rays by \textit{Fermi} within $\pm$3 days of the UVIT observations. Irrespective of their $\gamma$-ray brightness, the observed UV  brightening in both the sources may be attributed to fresh acceleration of electrons to higher energies as expected in  shock in-jet models \citep{marscher1985models,kirk1998particle}. Also, magnetic reconnection events or localized turbulence in the jet can lead to enhanced emission at shorter (bluer)  wavelengths  leading to the observed BWB trend. Changes in the Doppler factor due to geometrical effects, such as variations in the viewing angle of a curved or inhomogeneous jet could also lead to a BWB behaviour \citep{villata2004webt,papadakis2007long}.
\\[6pt]
A similar UV colour variability analysis by \cite{reshma2024ultraviolet} reported BWB trends in four BL Lacs and one FSRQ. This suggests that UV flux and colour variations observed in both FSRQs and BL Lacs are predominantly driven by synchrotron emission from relativistic jet electrons. Within the one zone leptonic emission model, a BWB trend can arise from the injection of freshly accelerated electrons with a harder energy distribution than the pre-existing cooler electron population \citep{kirk1998particle,mastichiadis2002models}.  Notably, the  two sources (one FSRQ and one BL Lac) in this study and the five sources reported by  \cite{reshma2024ultraviolet} that showed BWB in UV (that includes one FSRQ and four BL Lacs) are $\gamma$-ray loud sources detected by the \textit{Fermi}-LAT \citep{ackermann2015third}, further supporting the dominance of jet emission in these sources.

\section{Summary} 
\label{sec:summary}

The UV variability properties  of AGN remain relatively less explored compared to other wavelengths. Nevertheless, some studies do exist from HST \citep{dolan2004search}, IUE \citep{edelson1991rapid,sukanya2018long} and UVIT \citep{reshma2024ultraviolet} observations. In this work, we investigated UV variability on hour timescales using observations obtained with  UVIT. Our sample comprises 24 radio-loud (jetted) AGN, that includes 17 BL Lacs, 6 FSRQs and one radio-loud NLSy1 galaxy. The main results of this study are summarized below:
\\[3pt]
\begin{enumerate}
    \item [(i)] We detected UV variability in 18 out of the 24 sources, which includes the radio-loud NLSy1 galaxy 1H 0323+342, 5  FSRQs and 12 BL Lacs.  All the variable sources are jet dominated systems, strongly suggesting that the observed UV variability is driven by processes occurring within their relativistic jets. 
    \\[3pt]
    \item [(ii)] We found the  amplitude of variability, $\sigma_m$  in the NUV band to range from 0.010 mag to 0.046 mag, while in the FUV band, we found $\sigma_m$ to span a wider range between 0.010 mag to 0.247 mag. For the three sources with simultaneous FUV and NUV observations, the variability amplitude in FUV is larger than that in the  NUV band for CTA 102 and PKS 0447$-$439. This suggests that AGN may exhibit stronger variability at shorter UV wavelengths. However, a larger sample with simultaneous FUV and NUV coverage is required to confirm this trend robustly.
    \\[3pt]
    \item [(iii)] Constrained by the simultaneous observations in FUV and NUV bands, we investigated spectral variations in three sources, by generating colour magnitude diagrams. Of the three sources, significant colour variations were detected in two sources, namely, CTA 102 (FSRQ) and PKS 0447$-$439 (BL Lac). Both sources exhibited a clear BWB trend, which is most plausibly explained by the injection of fresh electrons with an energy distribution harder than the pre-existing cooler electrons in their jet.
\end{enumerate}
The observations reported here on the UV flux and spectral variations in our sample of $\gamma$-ray detected blazars favour a jet based origin of their variations. Future UV observations of a large number of blazars, particularly with simultaneous FUV and NUV observations will be crucial to establish whether both FSRQs and BL Lacs show a BWB behaviour in the UV band and if there is any potential connection between UV colour variability and the $\gamma$-ray brightness state of these sources.

\section*{Acknowledgements}
We sincerely thank the referee for the constructive comments and insightful suggestions, which have been invaluable in enhancing the quality and clarity of the manuscript. This research utilizes data from the \textit{AstroSat} mission led by the Indian Space Research Organization with the archival access provided by the Indian Space Science Data Centre (ISSDC). The UVIT data used in this work were processed by the Payload Operation Centre (POC) of the Indian Institute of Astrophysics (IIA), Bengaluru. The UVIT instrument was developed through a collaborative effort involving IIA, the Inter-University Centre for Astronomy and Astrophysics (IUCCA), the Tata Institute of Fundamental Research (TIFR), ISRO and the Canadian Space Agency (CSA). A.K.M. acknowledges the support from the European Research Council (ERC) under the European Union’s Horizon 2020 research and innovation program (grant No. 951549). One of the authors (SBG) thanks the Inter-University Centre for Astronomy and Astrophysics (IUCAA), Pune, India for the Visiting Associateship.

\bibliographystyle{elsarticle-harv} 
\bibliography{ref}

\appendix
\label{sec:appendix}

\setcounter{table}{0}
\section{Brightness measurements of the individual sources analyzed}
\begin{table*}[htbp]
\centering
\caption{Brightness measurements of 1ES 0120+340}
\begin{tabular}{ccccc}
&&&&\\
\hline
Date  & Filter & MJD start & MJD end & Mag \\
\hline
04-12-2018  & F154W  & 58453.177916 & 58453.184880 & 19.92  $\pm$ 0.14  \\
            &        & 58453.245573 & 58453.260578 & 19.64  $\pm$ 0.08  \\
            &        & 58453.313235 & 58453.333842 & 19.71  $\pm$ 0.07  \\
            &        & 58453.380897 & 58453.402049 & 19.65  $\pm$ 0.07  \\
            &        & 58453.448559 & 58453.469700 & 19.77  $\pm$ 0.07  \\
            &        & 58453.516222 & 58453.537350 & 19.78  $\pm$ 0.07  \\
\hline
\end{tabular}
\end{table*}

\begin{table*}
\centering
\caption{Brightness measurements of 1ES 0229+200}
\begin{tabular}{ccccc}
&&&&\\
\hline
Date  & Filter &  MJD start & MJD end & Mag  \\
\hline
22-09-2017  & F154W  & 58017.831844	& 58017.844789  & 20.50  $\pm$ 0.14   \\
            &        & 58017.897540	& 58017.914527  & 20.57  $\pm$ 0.12   \\
            &        & 58017.972337	& 58017.984271  & 20.41  $\pm$ 0.13   \\            
            &        & 58018.047132	& 58018.054009  & 20.47  $\pm$ 0.18   \\            
            &        & 58018.100982	& 58018.123759  & 20.66  $\pm$ 0.21   \\            
            &        & 58018.170720	& 58018.174450  & 20.45  $\pm$ 0.24   \\
\hline
\end{tabular}
\end{table*}            
            
\begin{table*}
\centering
\begin{tabular}{ccccc}
&&&&\\
\hline
Date  & Filter &  MJD start & MJD end & Mag  \\
\hline            
            &        & 59435.879428	& 59435.893830  & 20.43  $\pm$ 0.15  \\
            &        & 59435.948414	& 59435.961480  & 20.08  $\pm$ 0.13  \\
            &        & 59436.020983	& 59436.029142  & 19.98  $\pm$ 0.16  \\
            &        & 59436.420561	& 59436.435105  & 20.13  $\pm$ 0.13  \\
            &        & 59436.555851	& 59436.570429  & 20.26  $\pm$ 0.14  \\
            &        & 59436.623489	& 59436.638091  & 20.02  $\pm$ 0.12  \\
            &        & 59436.691133	& 59436.701052  & 20.27  $\pm$ 0.17  \\
            &        & 59436.758772	& 59436.773205  & 20.32  $\pm$ 0.14  \\
\hline
\end{tabular}
\end{table*}

\begin{table*}
\centering
\caption{Brightness measurements of AO 0235+164}
\begin{tabular}{ccccc}
&&&&\\
\hline
Date  & Filter &  MJD start & MJD end & Mag  \\
\hline
04-09-2016  & F148W  & 57634.342759	& 57634.350027 & 23.10 $\pm$ 0.75 \\           
            &        & 57634.413129	& 57634.421919 & 22.93 $\pm$ 0.60 \\
            &        & 57634.478083	& 57634.493996 & 23.36 $\pm$ 0.62 \\
            &        & 57634.545752	& 57634.561670 & 23.49 $\pm$ 0.68 \\
            &        & 57634.613414	& 57634.629344 & 22.77 $\pm$ 0.40 \\ 
            &        & 57634.681076	& 57634.697012 & 22.91 $\pm$ 0.44 \\
\hline
\end{tabular}
\end{table*}

\begin{table*}
\centering
\caption{Brightness measurements of 1H 0323+342}
\begin{tabular}{ccccc}
&&&&\\
\hline
Date  & Filter &  MJD start & MJD end & Mag  \\
\hline
06-09-2020  & F148W  & 59098.402173	& 59098.414745  & 17.93  $\pm$ 0.04  \\  
            &        & 59098.469829	& 59098.487177  & 17.98  $\pm$ 0.03  \\ 
            &        & 59098.537480	& 59098.554834  & 17.95  $\pm$ 0.03  \\
            &        & 59098.605136	& 59098.622490  & 17.99  $\pm$ 0.03  \\
            &        & 59098.672797	& 59098.690146  & 17.98  $\pm$ 0.03  \\
            &        & 59098.740443	& 59098.757547  & 17.95  $\pm$ 0.03  \\
30-09-2020  & F148W  & 59122.499163	& 59122.505618  & 17.84  $\pm$ 0.05  \\
            &        & 59122.552051	& 59122.573280  & 17.87  $\pm$ 0.03  \\
            &        & 59122.619695	& 59122.640937  & 17.86  $\pm$ 0.03  \\
            &        & 59122.687334	& 59122.708593  & 17.87  $\pm$ 0.03  \\
            &        & 59122.754973	& 59122.765409  & 17.92  $\pm$ 0.04  \\
04-02-2022  & F148W  & 59613.668914	& 59613.683080  & 17.86  $\pm$ 0.03  \\
            &        & 59613.736565	& 59613.750724  & 17.85  $\pm$ 0.03  \\
            &        & 59613.804235	& 59613.818363  & 17.81  $\pm$ 0.03  \\
            &        & 59613.875145	& 59613.885996  & 17.81  $\pm$ 0.04  \\
            &        & 59613.947714	& 59613.953635  & 17.90  $\pm$ 0.05  \\
            &        & 59614.142462	& 59614.150161  & 17.85  $\pm$ 0.04  \\
            &        & 59614.210113	& 59614.222731  & 17.82  $\pm$ 0.03  \\
            &        & 59614.277764	& 59614.291816  & 17.80  $\pm$ 0.03  \\
            &        & 59614.346997	& 59614.359449  & 17.83  $\pm$ 0.03  \\
            &        & 59614.413064	& 59614.422499  & 17.76  $\pm$ 0.04  \\ 
\hline
\end{tabular}
\end{table*}

\begin{table*}
\centering
\caption{Brightness measurements of 1ES 0347$-$121}
\begin{tabular}{ccccc}
&&&&\\
\hline
Date  & Filter &  MJD start & MJD end & Mag  \\
\hline
21-08-2020  & F148W  & 59082.304138 & 59082.312662 & 18.83 $\pm$ 0.06 \\
            &        & 59082.371788 & 59082.383845 & 19.00 $\pm$ 0.06 \\
            &        & 59082.439433 & 59082.450955 & 18.80 $\pm$ 0.06 \\
            &        & 59082.507077 & 59082.519524 & 18.82 $\pm$ 0.05 \\
\hline
\end{tabular}
\end{table*}

\begin{table*}
\centering
\caption{Brightness measurements of PKS 0352$-$686}
\begin{tabular}{ccccc}
&&&&\\
\hline
Date  & Filter &  MJD start & MJD end & Mag  \\
\hline
02-11-2018  & F148W  & 58423.476028	& 58423.494365  & 19.66  $\pm$ 0.07  \\
            &        & 58423.540074	& 58423.562033  & 19.76  $\pm$ 0.06  \\
            &        & 58423.607732	& 58423.629696  & 19.69  $\pm$ 0.06  \\
            &        & 58423.675394	& 58423.697357  & 19.73  $\pm$ 0.06  \\
            &        & 58423.743056	& 58423.765020  & 19.65  $\pm$ 0.06  \\
            &        & 58423.811259	& 58423.832683  & 19.67  $\pm$ 0.06  \\
            &        & 58423.883828	& 58423.896043  & 19.57  $\pm$ 0.08  \\
            & F154W  & 58423.956397	& 58423.967995  & 19.63  $\pm$ 0.09  \\
            &        & 58424.028966	& 58424.035657  & 19.56  $\pm$ 0.12  \\		
            &        & 58424.081368 & 58424.103308  & 19.66  $\pm$ 0.13  \\
            &        & 58424.149030	& 58424.157450  & 19.77  $\pm$ 0.12  \\
            &        & 58424.216693	& 58424.228627  & 19.55  $\pm$ 0.09  \\
            &        & 58424.284354	& 58424.299809  & 19.54  $\pm$ 0.08  \\
            &        & 58424.352023	& 58424.370645  & 19.54  $\pm$ 0.07  \\
            &        & 58424.419685	& 58424.441619  & 19.53  $\pm$ 0.07  \\	          
            &        & 58424.487347	& 58424.509282  & 19.64  $\pm$ 0.07  \\
            &        & 58424.555015	& 58424.569960  & 19.49  $\pm$ 0.08  \\         
\hline
\end{tabular}
\end{table*}

\begin{table*}
\centering
\caption{Brightness measurements of PKS 0447$-$439}
\begin{tabular}{ccccc}
&&&&\\
\hline
Date  & Filter &  MJD start & MJD end & Mag  \\
\hline
15-12-2017  & F154W  & 58102.012648	& 58102.021247  & 16.31  $\pm$ 0.02   \\
            &        & 58102.085217	& 58102.088921  & 16.30  $\pm$ 0.03   \\
            &        & 58102.134934	& 58102.140782  & 16.32  $\pm$ 0.03   \\
            &        & 58102.202597	& 58102.215087  & 16.30  $\pm$ 0.02   \\
            &        & 58102.270259	& 58102.287656  & 16.29  $\pm$ 0.02   \\
            &        & 58102.337922	& 58102.359535  & 16.31  $\pm$ 0.02   \\
            &        & 58102.405584	& 58102.427239  & 16.28  $\pm$ 0.02   \\
            &        & 58102.473252	& 58102.494907  & 16.29  $\pm$ 0.02   \\
            &        & 58102.540915	& 58102.544877  & 16.21  $\pm$ 0.03   \\
            & F169M  & 58102.546607	& 58102.562569  & 16.20  $\pm$ 0.02   \\
            & N245M  & 58102.012599	& 58102.021296  & 15.77  $\pm$ 0.01   \\    
            &        & 58102.085168	& 58102.088970  & 15.73  $\pm$ 0.02   \\
            &        & 58102.134886	& 58102.140831  & 15.78  $\pm$ 0.01   \\
            &        & 58102.202548	& 58102.215137  & 15.76  $\pm$ 0.01   \\
            &        & 58102.270210	& 58102.287706  & 15.75  $\pm$ 0.01   \\
            &        & 58102.337873	& 58102.359584  & 15.75  $\pm$ 0.01   \\
            &        & 58102.405535	& 58102.427288  & 15.76  $\pm$ 0.01   \\
            &        & 58102.473203	& 58102.494957  & 15.77  $\pm$ 0.01  \\
            &        & 58102.540866	& 58102.544927  & 15.75  $\pm$ 0.02  \\   
            & N263M  & 58102.546558	& 58102.562619  & 15.68  $\pm$ 0.01   \\ 	    
\hline
\end{tabular}
\end{table*}

\begin{table*}
\centering
\caption{Brightness measurements of 1H 0658+595}
\begin{tabular}{ccccc}
&&&&\\
\hline
Date  & Filter &  MJD start & MJD end & Mag  \\
\hline
19-11-2016  & F154W  & 57711.076889	& 57711.099403  & 18.99   $\pm$  0.09   \\
            &        & 57711.146694	& 57711.155726  & 18.96   $\pm$  0.08   \\
            &        & 57711.216492	& 57711.229507  & 18.88   $\pm$  0.06   \\
            &        & 57711.286291	& 57711.302930  & 18.92   $\pm$  0.05   \\
            &        & 57711.356095	& 57711.376347  & 18.98   $\pm$  0.05   \\
            & F172M  & 57711.440560	& 57711.446928  & 18.77   $\pm$  0.18    \\
            &        & 57711.498570	& 57711.518192  & 18.43   $\pm$  0.09   \\
            &        & 57711.565498	& 57711.587991  & 18.62   $\pm$  0.09   \\
            &        & 57711.635315	& 57711.657789  & 18.50   $\pm$  0.09   \\
            & N245M  & 57711.437081	& 57711.448446  & 18.50   $\pm$  0.04    \\
            &        & 57711.500361	& 57711.518244  & 18.51   $\pm$  0.03    \\
            &        & 57711.565446	& 57711.571027  & 18.45   $\pm$  0.05   \\
            & N263M  & 57711.577500 & 57711.588043  & 18.41   $\pm$  0.05   \\
            & N279N  & 57711.076839	& 57711.081639  & 18.39   $\pm$  0.15   \\
            &        & 57711.146643	& 57711.155777  & 18.41   $\pm$  0.11   \\
            &        & 57711.217441	& 57711.229558  & 18.43   $\pm$  0.09   \\
            &        & 57711.286240	& 57711.302981  & 18.29   $\pm$  0.08   \\
            &        & 57711.356045	& 57711.376398  & 18.39   $\pm$  0.07   \\
            &        & 57711.425843	& 57711.432454  & 18.35   $\pm$  0.13   \\
\hline
\end{tabular}
\end{table*}

\begin{table*}
\centering
\caption{Brightness measurements of S5 0836+71}
\begin{tabular}{ccccc}
&&&&\\
\hline
Date  & Filter &  MJD start & MJD end & Mag  \\
\hline
04-05-2017  & F154W  & 57875.583540	& 57875.605205 & 21.59  $\pm$ 0.19  \\
            &        & 57875.651212	& 57875.672878 & 21.32  $\pm$ 0.16  \\
            &        & 57875.718886	& 57875.740547 & 21.71  $\pm$ 0.20  \\
            &        & 57875.786554	& 57875.808215 & 21.71  $\pm$ 0.20  \\
            &        & 57875.857092	& 57875.875883 & 21.75  $\pm$ 0.22  \\
            &        & 57875.929661	& 57875.943545 & 21.57  $\pm$ 0.23  \\
\hline
\end{tabular}
\end{table*}

\begin{table*}
\centering
\caption{Brightness measurements of OJ 287}
\begin{tabular}{ccccc}
&&&&\\
\hline
Date  & Filter &  MJD start & MJD end & Mag  \\
\hline
25-10-2020  & F154W  & 59142.523770  & 59142.531976  & 17.19   $\pm$ 0.04    \\
            &        & 59142.597723  & 59142.599689  & 17.24   $\pm$ 0.07    \\
            &        & 59142.659065  & 59142.666768  & 17.24   $\pm$ 0.04    \\
            &        & 59142.934485  & 59142.937971  & 17.11   $\pm$ 0.05    \\ 
            &        & 59143.001761  & 59143.005620  & 17.23   $\pm$ 0.05    \\
            &        & 59143.069241  & 59143.073284  & 17.24   $\pm$ 0.05    \\
\hline
\end{tabular}
\end{table*}
            
\begin{table*}
\centering
\begin{tabular}{ccccc}
&&&&\\
\hline
Date  & Filter &  MJD start & MJD end & Mag  \\
\hline        
            &        & 59147.935560  & 59147.944724  & 17.41   $\pm$ 0.04    \\
            &        & 59148.003964  & 59148.012386  & 17.43   $\pm$ 0.04    \\
            &        & 59148.076533  & 59148.080036  & 17.38   $\pm$ 0.06    \\ 
            &        & 59148.409083  & 59148.415788  & 17.35   $\pm$  0.04   \\
            &        & 59148.476727  & 59148.485992  & 17.41   $\pm$  0.04   \\
            &        & 59148.544378  & 59148.552900  & 17.34   $\pm$  0.04   \\
\hline
\end{tabular}
\end{table*}

\begin{table*}
\centering
\begin{tabular}{ccccc}
&&&&\\
\hline
Date  & Filter &  MJD start & MJD end & Mag  \\
\hline  
20-11-2021  & F154W  & 59537.767828	& 59537.781827  & 17.47  $\pm$ 0.03  \\    
            &        & 59537.835472	& 59537.849478  & 17.45  $\pm$ 0.03  \\
            &        & 59537.903130	& 59537.911921  & 17.46  $\pm$ 0.04  \\
            &        & 59538.377030	& 59538.386623  & 17.47  $\pm$ 0.04  \\
            &        & 59538.444286	& 59538.458149  & 17.39  $\pm$ 0.03  \\
            &        & 59538.513955	& 59538.525995  & 17.44  $\pm$ 0.03  \\
\hline
\end{tabular}
\end{table*}

\begin{table*}
\centering
\caption{Brightness measurements of Mrk 180}
\begin{tabular}{ccccc}
&&&&\\
\hline
Date  & Filter &  MJD start & MJD end & Mag  \\
\hline
25-10-2018  & F148W  & 58415.633992	& 58415.645633 & 17.50   $\pm$ 0.03     \\
            &        & 58415.701625	& 58415.713295 & 17.51   $\pm$ 0.03     \\
            &        & 58415.769251	& 58415.780958 & 17.62   $\pm$ 0.03     \\
            &        & 58415.836884	& 58415.848620 & 17.52   $\pm$ 0.03     \\
            &        & 58415.904517	& 58415.916281 & 17.54   $\pm$ 0.03     \\
            &        & 58415.973063	& 58415.983939 & 17.56   $\pm$ 0.03     \\
            &        & 58416.242704	& 58416.248080 & 17.50   $\pm$ 0.04     \\
            &        & 58416.310336	& 58416.319606 & 17.53   $\pm$ 0.03     \\
            &        & 58416.377975	& 58416.389912 & 17.47   $\pm$ 0.03     \\
            &        & 58416.445613	& 58416.457574 & 17.55   $\pm$ 0.03     \\
            &        & 58416.513252	& 58416.525237 & 17.56   $\pm$ 0.03     \\
            &        & 58416.580885	& 58416.584151 & 17.47   $\pm$ 0.05     \\
            & F154W  & 58416.585873	& 58416.592893 & 17.49   $\pm$ 0.04    \\
            &        & 58416.648517	& 58416.660555 & 17.40   $\pm$ 0.03    \\
            &        & 58416.716150	& 58416.728218 & 17.43   $\pm$ 0.03    \\
            &        & 58416.783783	& 58416.795880 & 17.39   $\pm$ 0.03    \\
            &        & 58416.851421	& 58416.863542 & 17.45   $\pm$ 0.03    \\
\hline
\end{tabular}
\end{table*}

\begin{table*}
\centering
\caption{Brightness measurements of Ton 599}
\begin{tabular}{ccccc}
&&&&\\
\hline
Date  & Filter &  MJD start & MJD end & Mag  \\
\hline
24-04-2020  & F148W  & 58962.537295	& 58962.557688  &  19.29  $\pm$  0.05  \\
            &        & 58962.672613	& 58962.694573  &  19.25  $\pm$  0.05  \\
            &        & 58962.740270	& 58962.762224  &  19.17  $\pm$  0.05  \\
            &        & 58962.807940	& 58962.829892  &  19.18  $\pm$  0.05  \\
            &        & 58962.879655	& 58962.886370  &  19.08  $\pm$  0.08  \\
            & F169M  & 58962.888109	& 58962.897536  &  18.82  $\pm$  0.09  \\ 
            &        & 58962.952224	& 58962.965211  &  18.92  $\pm$  0.08   \\
            &        & 58963.024793	& 58963.032849  &  18.54  $\pm$  0.08  \\
            &        & 58963.097363	& 58963.100517  &  18.74  $\pm$  0.14  \\
            &        & 58963.146220	& 58963.150500  &  18.32  $\pm$  0.10  \\            
            &        & 58963.213876	& 58963.221333  &  18.62  $\pm$  0.09  \\            
            &        & 58963.281533	& 58963.292515  &  18.52  $\pm$  0.07  \\            
            &        & 58963.349195	& 58963.364041  &  18.41  $\pm$  0.06  \\            
            &        & 58963.416852	& 58963.435567  &  18.57  $\pm$  0.06  \\            
            &        & 58963.486657	& 58963.506485  &  18.38  $\pm$  0.05  \\            
            &        & 58963.552170	& 58963.557386  &  18.58  $\pm$  0.10  \\            
            & F172M  & 58963.559120	& 58963.574136  &  18.06  $\pm$  0.09   \\
            &        & 58963.619828	& 58963.641809  &  17.92  $\pm$  0.07   \\
            &        & 58963.687483	& 58963.709459  &  17.72  $\pm$  0.07   \\
            &        & 58963.755146	& 58963.777122  &  17.94  $\pm$  0.07   \\
            &        & 58963.823059	& 58963.844784  &  18.10  $\pm$  0.08   \\
            &        & 58963.895628	& 58963.905821  &  18.34  $\pm$  0.13   \\
\hline
\end{tabular}
\end{table*}

\begin{table*}
\centering
\caption{Brightness measurements of 4C +21.35}
\begin{tabular}{ccccc}
\hline
&&&&\\
Date  & Filter &  MJD start & MJD end & Mag  \\
\hline
25-04-2018  & F148W  & 58233.540491	& 58233.562165  & 16.32   $\pm$  0.02   \\
            &        & 58233.608182	& 58233.609904  & 16.40   $\pm$  0.04   \\
            & F172M  & 58233.611642	& 58233.629838  & 15.49   $\pm$  0.03   \\
  	    &   	 & 58233.675828	& 58233.697501  & 15.49   $\pm$  0.03   \\
  	  &        & 58233.743496 & 58233.765164  & 15.49   $\pm$  0.03   \\
    	&   	 & 58233.811172	& 58233.832820  & 15.49   $\pm$  0.03   \\
    	&   	 & 58233.883480	& 58233.891753  & 15.48   $\pm$  0.04   \\     
\hline
\end{tabular}
\end{table*}

\begin{table*}
\centering
\caption{Brightness measurements of OQ 334}
\begin{tabular}{ccccc}
&&&&\\
\hline
Date  & Filter &  MJD start & MJD end & Mag  \\
\hline
12-01-2020  & F148W  & 58859.704781	& 58859.715569  & 16.37 $\pm$ 0.02  \\
            &        & 58859.772426	& 58859.783237  & 16.37 $\pm$ 0.02  \\
            &        & 58859.840070	& 58859.850893  & 16.13 $\pm$ 0.02  \\
            &        & 58859.907714	& 58859.918550  & 16.36 $\pm$ 0.02  \\
            &        & 58859.975378	& 58859.979071  & 16.44 $\pm$ 0.03  \\
            & F154W  & 58859.980809	& 58859.986206  & 16.16 $\pm$ 0.03  \\
            &        & 58860.047023	& 58860.052696  & 16.25 $\pm$ 0.03  \\
            &        & 58860.119592	& 58860.121525  & 16.26 $\pm$ 0.05  \\            
            &        & 58860.245949	& 58860.249121  & 16.14 $\pm$ 0.03  \\            
            &        & 58860.313599	& 58860.320648  & 16.33 $\pm$ 0.03  \\            
            &        & 58860.381244	& 58860.392162  & 16.45 $\pm$ 0.02  \\            
            &        & 58860.448894	& 58860.459818  & 16.61 $\pm$ 0.02  \\            
            & F169M  & 58860.519803	& 58860.527475  & 16.36 $\pm$ 0.03  \\
            &        & 58860.584189	& 58860.595137  & 16.59 $\pm$ 0.03  \\            
            &        & 58860.651834	& 58860.662799  & 16.63 $\pm$ 0.03  \\            
            &        & 58860.719484	& 58860.730455  & 16.56 $\pm$ 0.03  \\            
            &        & 58860.787129	& 58860.793514  & 16.53 $\pm$ 0.04  \\            
            & F172M  & 58860.795252	& 58860.798112  & 16.05 $\pm$ 0.08  \\ 
\hline
\end{tabular}
\end{table*}

\begin{table*}
\centering
\caption{Brightness measurements of Mrk 501}
\begin{tabular}{ccccc}
&&&&\\
\hline
Date  & Filter &  MJD start & MJD end & Mag  \\
\hline
25-03-2022  & F154W  & 59662.314522	& 59662.326899  & 15.88  $\pm$ 0.02   \\
            &        & 59662.381648	& 59662.397732  & 15.87  $\pm$ 0.02   \\
            &        & 59662.449286	& 59662.466911  & 15.90  $\pm$ 0.02   \\
            &        & 59662.516919	& 59662.534562  & 15.89  $\pm$ 0.02   \\
            &        & 59662.584552	& 59662.602212  & 15.89  $\pm$ 0.02   \\
            &        & 59662.652190	& 59662.669863  & 15.90  $\pm$ 0.02   \\
\hline
\end{tabular}
\end{table*}

\begin{table*}
\centering
\begin{tabular}{ccccc}
&&&&\\
\hline
Date  & Filter &  MJD start & MJD end & Mag  \\
\hline  
            &        & 59874.168450	& 59874.173239  & 15.85  $\pm$ 0.02  \\
            &        & 59874.236088	& 59874.240866  & 15.81  $\pm$ 0.02  \\
            &        & 59874.303721	& 59874.308505  & 15.84  $\pm$ 0.02  \\
            &        & 59874.372252	& 59874.376131  & 15.87  $\pm$ 0.02  \\
            &        & 59874.438992	& 59874.443770  & 15.80  $\pm$ 0.02  \\
            &        & 59874.506625	& 59874.511415  & 15.83  $\pm$ 0.02  \\
\hline
\end{tabular}
\end{table*}

\begin{table*}
\centering
\caption{Brightness measurements of 1H 1720+117}
\begin{tabular}{ccccc}
&&&&\\
\hline
Date  & Filter &  MJD start & MJD end & Mag  \\
\hline
10-05-2018  & F148W  & 58248.765601 & 58248.787174 	& 18.50   $\pm$ 0.04    \\
            &        & 58248.833828 & 58248.845706 	& 18.51   $\pm$ 0.05    \\
\hline
\end{tabular}
\end{table*}

\begin{table*}
\centering
\caption{Brightness measurements of S3 1741+19}
\begin{tabular}{ccccc}
&&&&\\
\hline
Date  & Filter &  MJD start & MJD end & Mag  \\
\hline
23-08-2019  & F154W  & 58717.672011	& 58717.689513  & 19.36  $\pm$  0.07     \\
            &        & 58717.739674	& 58717.757170	& 19.55  $\pm$  0.07    \\
            &        & 58717.807337	& 58717.824820  & 19.54  $\pm$  0.07    \\
            &        & 58717.879314	& 58717.892471  & 19.50  $\pm$  0.08    \\
            &        & 58717.951883	& 58717.960127  & 19.50  $\pm$  0.10     \\
            &        & 58718.024452	& 58718.027765  & 19.28  $\pm$  0.15     \\
            &        & 58718.213310	& 58718.223420  & 19.38  $\pm$  0.09     \\
            &        & 58718.280972	& 58718.287319  & 19.37  $\pm$  0.11     \\
            & F172M  & 58718.289059	& 58718.295639  & 18.71  $\pm$  0.17    \\
            &        & 58718.348634	& 58718.366017  & 18.82  $\pm$  0.11    \\
            &        & 58718.416296	& 58718.433680  & 19.03  $\pm$  0.12    \\
            &        & 58718.483959	& 58718.501330  & 19.05  $\pm$  0.12    \\
            &        & 58718.551627	& 58718.568980  & 18.84  $\pm$  0.11    \\
            &        & 58718.619289	& 58718.636631  & 18.90  $\pm$  0.12    \\
            &        & 58718.754614	& 58718.771932  & 19.06  $\pm$  0.12    \\
            &        & 58718.822716	& 58718.830209  & 18.44  $\pm$  0.14    \\
\hline
\end{tabular}
\end{table*}

\begin{table*}
\centering
\caption{Brightness measurements of 1ES 1959+650}
\begin{tabular}{ccccc}
&&&&\\
\hline
Date  & Filter &  MJD start & MJD end & Mag  \\
\hline
05-10-2016  & F154W  & 57666.211111	& 57666.223330  & 16.29   $\pm$  0.02      \\
            &        & 57666.280721	& 57666.296546  & 16.27   $\pm$  0.02      \\
            &        & 57666.350324	& 57666.358387  & 16.28   $\pm$  0.02      \\
            & N263M  & 57666.211060	& 57666.223381 	& 15.54   $\pm$  0.01     \\
            &        & 57666.280670	& 57666.296596  & 15.53   $\pm$  0.01      \\
            &        & 57666.350274	& 57666.358438  & 15.54   $\pm$  0.02      \\
\hline
\end{tabular}
\end{table*}

\begin{table*}
\centering
\begin{tabular}{ccccc}
&&&&\\
\hline
Date  & Filter &  MJD start & MJD end & Mag  \\
\hline  
25-10-2017  & F154W  & 58051.181646	& 58051.186616  & 15.99  $\pm$  0.03  \\
            &        & 58051.249315	& 58051.263010  & 15.99  $\pm$  0.02    \\
            &        & 58051.316977	& 58051.334720  & 15.99  $\pm$  0.02    \\
            &        & 58051.384645	& 58051.402394  & 15.99  $\pm$  0.02    \\
            &        & 58051.452314	& 58051.470080  & 15.96  $\pm$  0.02    \\
            &        & 58051.519976	& 58051.537754  & 15.94  $\pm$  0.02    \\
            &        & 58051.587650	& 58051.605441  & 15.94  $\pm$  0.02    \\
            &        & 58051.655312	& 58051.673115  & 15.96  $\pm$  0.02    \\
            &        & 58051.722974	& 58051.740795  & 15.94  $\pm$  0.02    \\
            &        & 58051.790646	& 58051.808469  & 15.94  $\pm$  0.02    \\
            &        & 58051.861607	& 58051.875679  & 15.94  $\pm$  0.02    \\            
            & N279N  & 58051.181597	& 58051.186666  & 15.35  $\pm$  0.03    \\    
            &        & 58051.249266	& 58051.263060  & 15.25  $\pm$  0.02    \\
            &        & 58051.316928	& 58051.334770  & 15.24  $\pm$  0.02    \\
            &        & 58051.384596	& 58051.402444  & 15.27  $\pm$  0.02    \\
            &        & 58051.452264	& 58051.470130  & 15.21  $\pm$  0.02    \\
            &        & 58051.519927	& 58051.537804  & 15.23  $\pm$  0.02    \\
            &        & 58051.587601	& 58051.605490  & 15.18  $\pm$  0.02    \\
            &        & 58051.655263	& 58051.673164  & 15.22  $\pm$  0.02    \\
            &        & 58051.722925	& 58051.740844  & 15.22  $\pm$  0.02    \\
            &        & 58051.790577	& 58051.808519  & 15.19  $\pm$  0.02    \\
            &        & 58051.861558	& 58051.875729  & 15.17  $\pm$  0.02    \\
\hline
\end{tabular}
\end{table*}

\begin{table*}
\centering
\caption{Brightness measurements of PKS 2005$-$489}
\begin{tabular}{ccccc}
&&&&\\
\hline
Date  & Filter &  MJD start & MJD end & Mag  \\
\hline
15-10-2019  & F148W  & 58771.266239	& 58771.266236  & 16.38  $\pm$ 0.02     \\
            &        & 58771.332523	& 58771.331952  & 16.40  $\pm$ 0.02     \\
            &        & 58771.394770	& 58771.409334  & 16.39  $\pm$ 0.02     \\
            &        & 58771.462426	& 58771.465491  & 16.40  $\pm$ 0.03     \\
            & F154W  & 58771.467225	& 58771.476988  & 16.27  $\pm$ 0.02     \\
            &        & 58771.530088	& 58771.544651  & 16.26  $\pm$ 0.02     \\
            &        & 58771.597767	& 58771.612301  & 16.27  $\pm$ 0.02     \\
            &        & 58771.665407	& 58771.673344  & 16.26  $\pm$ 0.02     \\
            & F169M  & 58771.675082	& 58771.679952	& 16.18  $\pm$ 0.03      \\
            &        & 58771.733069	& 58771.747602  & 16.25  $\pm$ 0.02     \\
            &        & 58771.800739	& 58771.815253  & 16.21  $\pm$ 0.02      \\
            &        & 58771.872715	& 58771.881982  & 16.20  $\pm$ 0.03     \\ 
\hline
\end{tabular}
\end{table*}

\begin{table*}
\centering
\caption{Brightness measurements of CTA 102}
\begin{tabular}{ccccc}
&&&&\\
\hline
Date  & Filter &  MJD start & MJD end & Mag  \\
\hline
21-07-2017  & F148W  & 57954.350649	& 57954.366961 	& 18.67  $\pm$  0.04     \\
            &        & 57954.423957	& 57954.442245  & 18.68  $\pm$  0.04    \\ 
            &        & 57954.489820	& 57954.511835 	& 18.76  $\pm$  0.04    \\
            &        & 57954.559405	& 57954.581426 	& 18.70  $\pm$  0.04    \\	
            &        & 57954.636480	& 57954.651030 	& 18.69  $\pm$  0.05    \\
            &        & 57954.698568	& 57954.720615 	& 18.70  $\pm$  0.04    \\
\hline
\end{tabular}
\end{table*}

\begin{table*}
\centering
\caption{Brightness measurements of 3C 454.3}
\begin{tabular}{ccccc}
&&&&\\
\hline
Date  & Filter &  MJD start & MJD end & Mag  \\
\hline
20-10-2016  & F154W  & 57680.625139	& 57680.646044  & 18.26  $\pm$ 0.04    \\
            &        & 57680.694799	& 57680.715692  & 18.31  $\pm$ 0.04    \\
            &        & 57680.764458	& 57680.785339  & 18.35  $\pm$ 0.04     \\
            &        & 57680.835828	& 57680.854987  & 18.28  $\pm$ 0.04     \\
            &        & 57680.910526	& 57680.924606	& 18.30  $\pm$ 0.04     \\
            &        & 57680.985224	& 57680.994276	& 18.29  $\pm$ 0.06     \\
\hline
\end{tabular}
\end{table*}

\begin{table*}
\centering
\caption{Brightness measurements of 1ES 2322$-$409}
\begin{tabular}{ccccc}
&&&&\\
\hline
Date  & Filter &  MJD start & MJD end & Mag  \\
\hline
03-07-2020  & F148W  & 59033.450076	& 59033.465859  & 17.76  $\pm$  0.03     \\
            &        & 59033.517732	& 59033.529611  & 17.77  $\pm$  0.03     \\ 
            &        & 59033.585388	& 59033.601190  & 17.74  $\pm$  0.03     \\
            &        & 59033.654843	& 59033.664081  & 17.71  $\pm$  0.04     \\  
\hline
\end{tabular}
\end{table*}

\begin{table*}
\centering
\caption{Brightness measurements of 1ES 2344+514}
\begin{tabular}{ccccc}
&&&&\\
\hline
Date  & Filter &  MJD start & MJD end & Mag  \\
\hline
22-10-2017  & F172M  & 58048.015081	& 58048.021983  & 19.09   $\pm$ 0.19  \\
            &        & 58048.207274	& 58048.209186  & 18.89   $\pm$ 0.33  \\
            &        & 58048.272010	& 58048.285228  & 19.13   $\pm$ 0.14  \\
            & N245M  & 58047.952626	& 58047.955603  & 18.86   $\pm$ 0.09  \\  
            &        & 58048.015032	& 58048.017860  & 18.95   $\pm$ 0.10  \\
            &        & 58048.207274	& 58048.209236  & 19.01   $\pm$ 0.12  \\
\hline
\end{tabular}
\end{table*}

\begin{table*}
\centering
\caption{Brightness measurements of H 2356$-$309}
\begin{tabular}{ccccc}
&&&&\\
\hline
Date  & Filter &  MJD start & MJD end & Mag  \\
\hline
17-10-2018  & F154W  & 58407.368693	& 58407.390752  & 18.31  $\pm$ 0.04  \\
            &        & 58407.504017	& 58407.526088  & 18.23  $\pm$ 0.04  \\
            &        & 58407.571680	& 58407.593751  & 18.28  $\pm$ 0.04  \\
            &        & 58407.639342	& 58407.661413  & 18.22  $\pm$ 0.04  \\
            &        & 58407.707005	& 58407.729075  & 18.30  $\pm$ 0.04  \\
            &        & 58407.774667	& 58407.796731  & 18.23  $\pm$ 0.04  \\
\hline
\end{tabular}
\end{table*}

\end{document}